\documentclass[fleqn,useAMS,usenatbib]{mnras}
\usepackage{mathptmx}
\usepackage{color}
\usepackage{graphics,graphicx,amssymb,amsmath,natbib}
\usepackage{float}
\usepackage{color}
\usepackage{lscape,graphicx}
\usepackage{amsmath}
\usepackage{amssymb}
\usepackage{multirow}
\usepackage{afterpage}
\usepackage{threeparttable}
\usepackage{subfigure}
\usepackage{rotating}
\usepackage{lscape}
\usepackage{caption}
\usepackage{verbatim}
\usepackage{morefloats}
\usepackage[T1]{fontenc}
\usepackage{ae,aecompl}
\usepackage{graphicx}
\usepackage{epstopdf}
\usepackage{amsmath}
\usepackage{amssymb}
\usepackage{enumerate}
\usepackage{dblfloatfix}
\usepackage{lipsum}

\def\cha    {{\em Chandra}\/}

\def\xmm        {{\em XMM-Newton}\/}
\def\XMM        {{\em XMM}\/}

\def\ga         {{CGCG~254-021}\/}

\def\sdss        {{\em SDSS}\/}
\def\subaru        {{\em Subaru}\/}
\def\muse        {{\em MUSE}\/}

\newcommand{\hi}{H\,{\sc i}}
\newcommand{\hii}{H\,{\sc ii}}
\newcommand{\km}{km\,s$^{-1}$}
\newcommand{\OIII}{\mbox{[O\,\textsc{iii}]}}
\newcommand{\OI}{\mbox{[O\,\textsc{i}]}}
\newcommand{\NII}{\mbox{[N\,\textsc{ii}]}}
\newcommand{\SII}{\mbox{[S\,\textsc{ii}]}}
\newcommand{\RNum}[1]{\uppercase\expandafter{\romannumeral #1\relax}}

\title[An H$\alpha$/X-ray orphan cloud]{An H$\alpha$/X-ray orphan cloud as a signpost of intracluster medium clumping}
\author
[Ge et al.]{
Chong Ge$^{1}$\thanks{chong.ge@uah.edu}, 
Rongxin Luo$^{1}$, 
Ming Sun$^{1}$\thanks{ming.sun@uah.edu}, 
Masafumi Yagi$^{2}$,
Pavel J\'{a}chym$^{3}$,
\newauthor
Alessandro Boselli$^{4}$,
Matteo Fossati$^{5}$,
Paul E.J. Nulsen$^{6}$,
Craig Sarazin$^{7}$, 
\newauthor
Tim Edge$^{1}$,
Giuseppe Gavazzi$^{5}$,
Massimo Gaspari$^{8,9}$,
Jin Koda$^{10}$, 
\newauthor
Yutaka Komiyama$^{2}$,
and Michitoshi Yoshida$^{11}$ \\
$^{1}$Department of Physics and Astronomy, University of Alabama in Huntsville, Huntsville, AL 35899, USA\\
$^{2}$National Astronomical Observatory of Japan, 2-21-1, Osawa, Mitaka, Tokyo, 181-8588, Japan\\
$^{3}$Astronomical Institute, Academy of Sciences of the Czech Republic, Bo\v{c}n\'{l} II 1401, 14100 Prague, Czech Republic\\
$^{4}$Aix Marseille Universit\'{e}, CNRS, LAM (Laboratoire d’ Astro physique de Marseille) UMR 7326, 13388, Marseille, France\\ 
$^{5}$Dipartimento di Fisica G. Occhialini, Universit\'{a} degli Studi di Milano Bicocca, Piazza della Scienza 3, I-20126 Milano, Italy\\
$^{6}$Center for Astrophysics \textbar{} Harvard \& Smithsonian, Cambridge, MA 02138, USA\\
$^{7}$Department of Astronomy, University of Virginia, P.O. Box 400325, Charlottesville, VA 22901-4325, USA\\
$^{8}$INAF, Osservatorio di Astrofisica e Scienza dello Spazio, via P. Gobetti 93/3, 40129 Bologna, Italy\\
$^{9}$Department of Astrophysical Sciences, Princeton University, 4 Ivy Lane, Princeton, NJ 08544, USA\\
$^{10}$Department of Physics and Astronomy, Stony Brook University 100 Nicolls Rd., Stony Brook, NY 11794-3800, USA\\
$^{11}$Subaru Telescope, National Astronomical Observatory of Japan, 650 North A'ohoku Place, Hilo, HI 96720, USA\\
}
\begin{document}
\date{Accepted. Received; in original form}

\pubyear{2021}

\maketitle
\begin{abstract}
Recent studies have highlighted the potential significance of intracluster medium (ICM) clumping and its important implications for cluster cosmology and baryon physics.
Many of the ICM clumps can originate from infalling galaxies, as stripped interstellar medium (ISM) mixing into the hot ICM.
However, a direct connection between ICM clumping and stripped ISM has not been unambiguously established before.
Here we present the discovery of the first and still the only known isolated cloud (or orphan cloud; OC) detected in both X-rays and H$\alpha$ in the nearby cluster A1367.
With an effective radius of 30 kpc, this cloud has an average X-ray temperature of 1.6 keV, a bolometric X-ray luminosity of $\sim 3.1\times 10^{41} {\rm\ erg\ s}^{-1}$ and a hot gas mass of $\sim 10^{10}\ {\rm M}_\odot$.
From the \muse\ data, the OC shows an interesting velocity gradient nearly along the east-west direction with a low level of velocity dispersion of $\sim 80$ \km, which may suggest a low level of the ICM turbulence. The emission line diagnostics suggest little star formation in the main H$\alpha$ cloud and a LI(N)ER-like spectrum, but the excitation mechanisms remain unclear. This example shows that stripped ISM, even long after the initial removal from the galaxy, can still induce ICM inhomogeneities. We suggest that the magnetic field can stabilize the OC by suppressing hydrodynamic instabilities and thermal conduction.
This example also suggests that at least some ICM clumps are multi-phase in nature
and implies that the ICM clumps can also be traced in H$\alpha$. Thus, future deep and wide-field H$\alpha$ surveys can be used to probe the ICM clumping and turbulence.
\end{abstract}

\begin{keywords}
galaxies: clusters: individual: Abell 1367 -- galaxies: clusters: intracluster medium -- galaxies: ISM -- X-rays: galaxies: clusters 
\end{keywords}
 
\section{Introduction} \label{sec:intro}
Galaxy clusters grow hierarchically through merging and the accretion of smaller structures along the cosmic filaments, which are continuously channeling dark matter, galaxies, and gas into clusters. 
As galaxies enter the cluster environment filled with hot intracluster medium (ICM) with $T \sim 10^{7} - 10^{8}$ K, their interstellar medium (ISM) is depleted by ram pressure and turbulent/viscous stripping from ICM (e.g. \citealt{1972ApJ...176....1G,2000Sci...288.1617Q}). These stripping processes are very important to the evolution of the cluster galaxies through rapidly quenching their star formation (SF) activities, and eventually may turn blue disk galaxies into red galaxies (e.g. \citealt{2006PASP..118..517B}). The stripping tails of cluster late-type galaxies have been observed from radio, mm, IR, and optical to X-ray (e.g. \citealt{2001ApJ...563L..23G,2002ApJ...567..118Y,2007ApJ...660.1209Y,2007ApJ...659L.115C,2008ApJ...687L..69K,2010ApJ...708..946S,2010ApJ...717..147S,2013MNRAS.429.1747M,2014ApJ...792...11J,2016A&A...587A..68B,2020MNRAS.496.4654C}).
In contrast to the early general wisdom that the stripped cold gas will simply mix with the hot ICM and be heated, now it is known that some fraction of the stripped ISM can collapse and form stars in the galactic halo and the intracluster space, especially in high-ICM-pressure environments (e.g. \citealt{2007ApJ...671..190S,2008ApJ...688..918Y,2010MNRAS.408.1417S,2013ApJ...778...91Y,2016AJ....151...78P}). 

Apart from the stripped tails close to their host galaxies, recent \hi\ surveys have also revealed the existence of a population of optically dark, isolated \hi\ clouds in galaxy clusters (e.g. \citealt{2004MNRAS.349..922D,2007ApJ...665L..15K}). 
The typical cloud mass is $\sim10^7\ {\rm M}_\odot$ with a size around a few kpc (e.g. \citealt{2012MNRAS.423..787T,2016ApJ...824L...7B}).
Despite the initial excitement for the so-called `dark galaxies', follow-up studies (e.g. \citealt{2008ApJ...673..787D,2016MNRAS.461.3001T}) suggest that these isolated clouds are most likely debris of ram pressure stripping (RPS) and tidal interaction. 

Around the same time, ICM clumping has been revealed in the X-ray data (e.g. \citealt{2011Sci...331.1576S,2011ApJ...731L..10N,2012MNRAS.421.1123C,2015MNRAS.447.2198E,2017MNRAS.469.2423M}). Many of the X-ray clumps are likely evaporating cold gas removed from galaxies (e.g \citealt{2009MNRAS.399..497D,2013MNRAS.429..799V}). The stripped gas clouds induce inhomogeneity or clumpiness in the ICM. Since the X-ray emissivity of the ICM scales with the square of gas density, ICM clumpiness can bias the measured gas density, which will further bias the gas mass, entropy, pressure, and cluster mass (e.g. \citealt{2011Sci...331.1576S,2011ApJ...731L..10N,2013MNRAS.429..799V}). 
In addition to clumpiness, turbulence in the ICM provides additional pressure against gravity, thus it can also bias the mass determinations assuming hydrostatic equilibrium if it is not accounted for (e.g. \citealt{2009ApJ...705.1129L}).
The characterization of ICM clumpiness and turbulence is important for current and next-generation surveys in the X-ray and millimeter via the Sunyaev-Zel’dovich effect, as well as using clusters as precise cosmological probes. However, there is limited information about the properties of individual ICM clumps from both observations and simulations.

\begin{figure}
\begin{center}
\centering
\includegraphics[angle=0,width=0.48\textwidth]{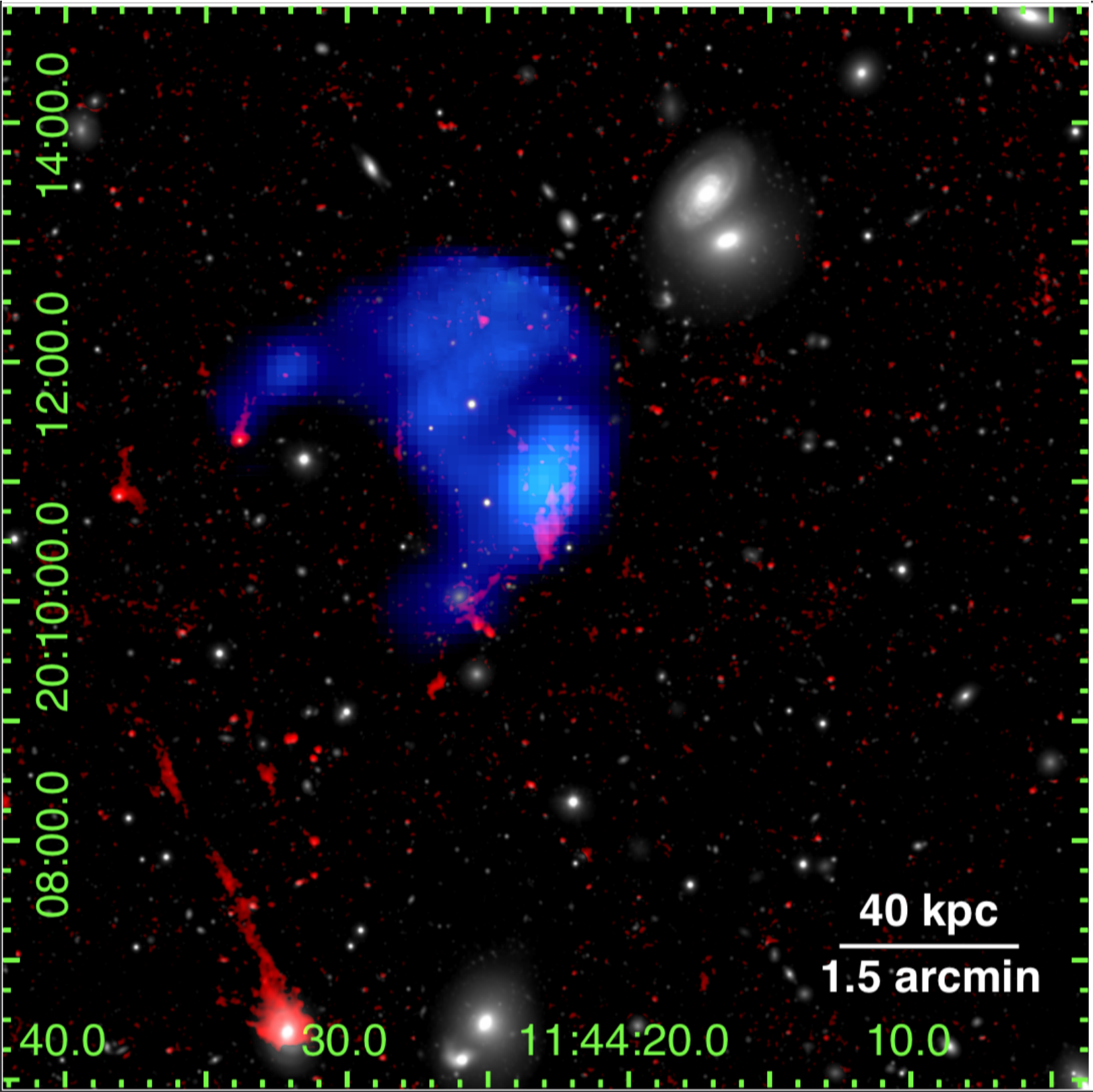}
\caption{
Three-color composite X-ray/optical image around the orphan cloud (OC). White: \subaru\ $r$-band image; red: \subaru\ net H$\alpha$ image; blue: \XMM\ 0.5-2 keV image. The RA and Dec are in J2000.
}
\label{fig:rgb}
\end{center}
\end{figure}

We recently discovered an isolated X-ray clump with a counterpart in the form of warm ionized gas in the nearby galaxy cluster A1367, which is a dynamically unrelaxed cluster in the Coma supercluster (e.g. \citealt{2002ApJ...576..708S}; \citealt{2004A&A...425..429C}). This cloud was first discovered in a narrow-band H$\alpha$ imaging survey of A1367 \citep{2017ApJ...839...65Y}. However, its velocity was unknown so its origin remained unclear. It was classified as an orphan cloud (OC; Fig.~\ref{fig:rgb}). Our follow-up {\em XMM} observation in this field to study cluster merger shock and X-ray tails \citep{2019MNRAS.486L..36G} unexpectedly revealed a diffuse soft X-ray clump around the same position as the H$\alpha$ OC (Fig.~\ref{fig:rgb}). Finally, our new {\em MUSE} data confirm its association with A1367.
The A1367 OC presents a great laboratory to study the evolution of the stripped ISM far away from the parent galaxy, and meanwhile to study the ICM clumping in detail.
Here we present a multi-wavelength study for this isolated (or galaxy-less) cloud.
We assume a cosmology with $H_0$ = 70 km s$^{-1}$ Mpc$^{-1}$, $\Omega_m=0.3$, and $\Omega_{\Lambda}= 0.7$. At A1367's redshift of $z=0.022$, $1^{\prime\prime}=0.445$ kpc.

\begin{figure*}
\begin{center}
\centering
\includegraphics[angle=0,width=0.99\textwidth]{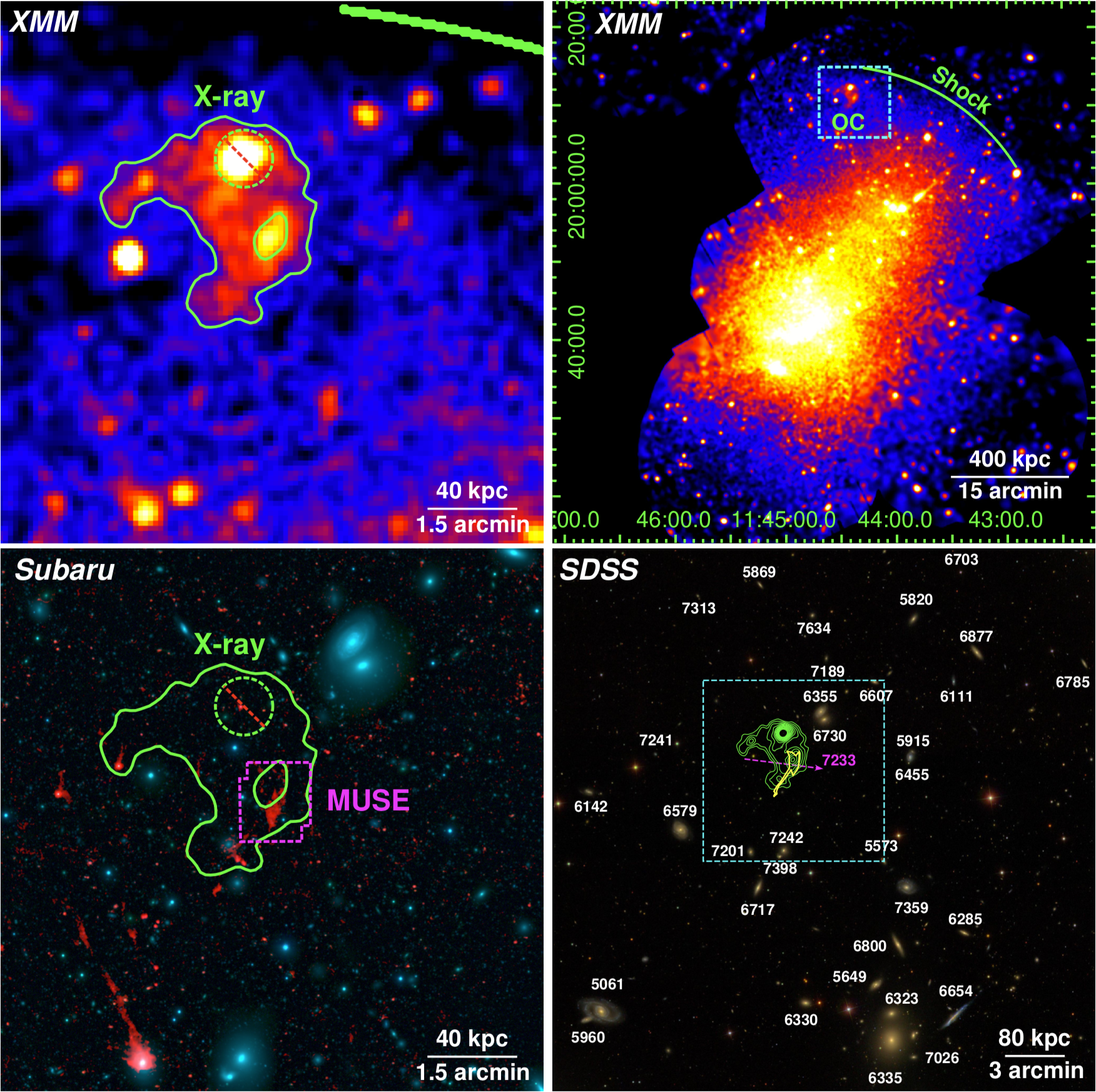}
\caption{
X-ray and optical images of the OC and nearby galaxies.
\textit{\textbf{Upper left:}} the 0.5-2 keV \XMM\ image of the OC, background subtracted and exposure corrected. The green contour outlines its X-ray morphology and the dashed circle marks a bright point source that is most likely an unrelated background AGN.
\textit{\textbf{Upper right:}} the 0.5-2 keV \XMM\ mosaic of A1367 (background subtracted and exposure corrected), with the dashed cyan box showing the field of the left panels. The OC is marked and the green arc marks a merger shock front \protect\citep{2019MNRAS.486L..36G}. Part of the green arc is shown in the upper left panel. 
\textit{\textbf{Lower left:}} the \subaru\ three-color composite image (red: net H$\alpha$; green: $r$-band; blue: $g$-band) of the same field as the upper left panel.
The dashed magenta region shows the \muse\ FOV. 
\textit{\textbf{Lower right:}} \sdss\ image around the OC, with the green contours from X-rays. The yellow contour highlights the H$\alpha$ cloud from the \subaru\ image \protect\citep{2017ApJ...839...65Y}. The velocities of galaxies in A1367 are marked with white numbers (in a unit of \km). The velocity of the OC is marked with a magenta number. The direction of the velocity gradient is marked with a dashed magenta arrow (see Fig.~\ref{fig:muse_maps} below).
The dashed cyan box shows the field of the left panels.
}
\label{fig:img}
\end{center}
\end{figure*}

\begin{figure*}
\begin{center}
\includegraphics[angle=0,width=1\textwidth]{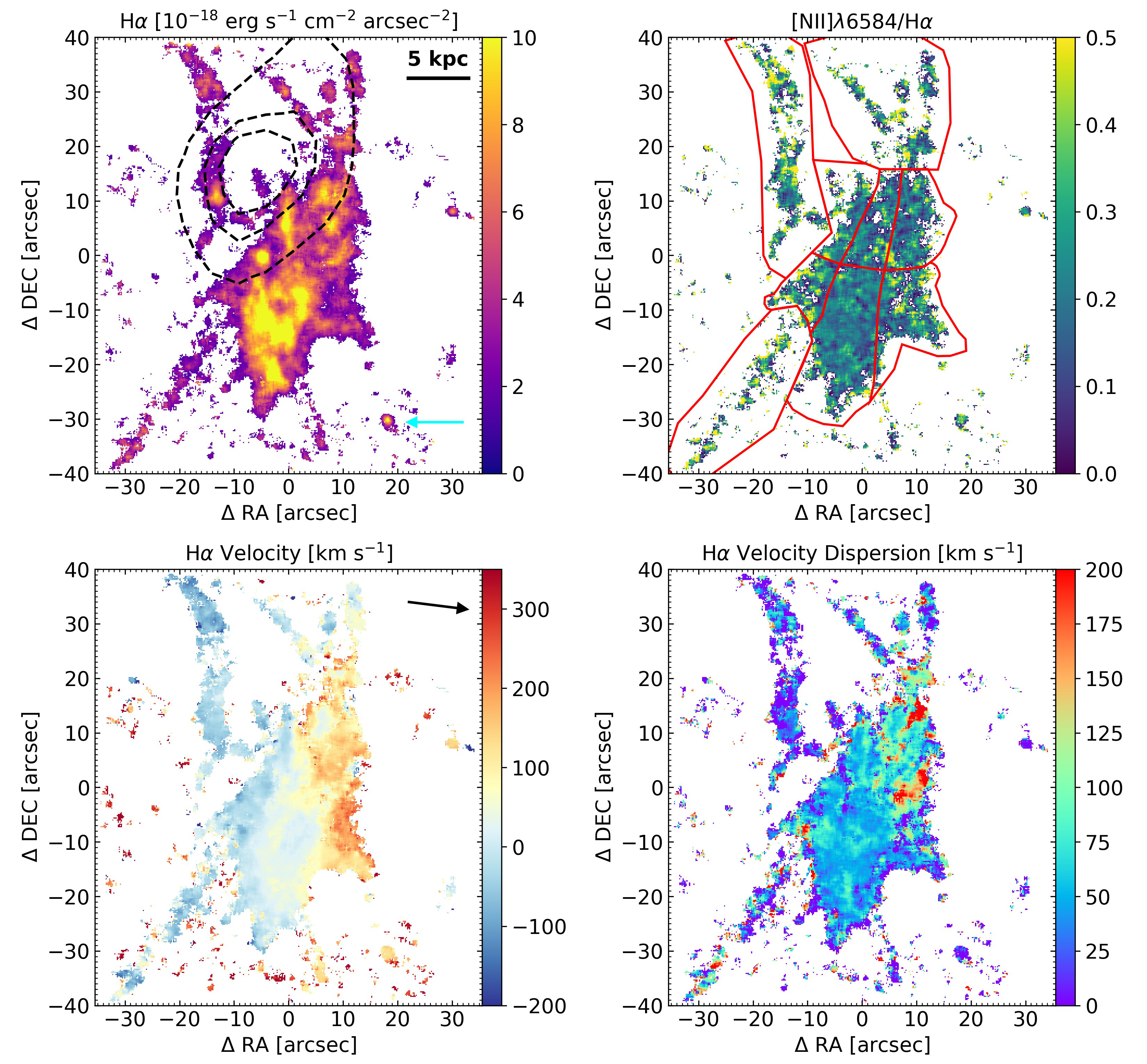}
\caption{The two-dimensional maps on the properties of the warm, ionized gas in the OC from the \muse\ observations, relative to (11:44:22.74, +20:10:44.60).
\textit{\textbf{Upper left:}} H$\alpha$ surface brightness. The black contours in the dashed lines show the central X-ray emission of the OC, with the outer contour the same as the one around the X-ray peak in the left panels of Fig.~\ref{fig:img}. The cyan arrow shows the only candidate \hii\ region in the {\em MUSE} field.
\textit{\textbf{Upper right:}} \NII/H$\alpha$ flux ratio. The red solid lines show the 9 large regions where the total spectra are extracted for studies of line diagnostics and kinematics.
\textit{\textbf{Lower left:}} H$\alpha$ velocity (relative to $z=0.024$). The black arrow shows the best-fit direction of the velocity gradient.
\textit{\textbf{Lower right:}} H$\alpha$ velocity dispersion.
}
\label{fig:muse_maps}
\end{center}
\end{figure*}

\section{Data analysis}
\label{s:obs}
 \subsection{\xmm\ data processing}
We analyzed the data with the Obsid of 0823200101 (PI: M. Sun; total time: 71.6 ks; clean time: 66.8 ks for MOS and 50.4 ks for pn) for the properties of OC. The mosaic image of A1367 is from our previous study \citep{2019MNRAS.486L..36G}, updated with a new observation with Obsid of 0864410101 (PI: C. Ge; total time: 40.0 ks; clean time: 32.8 ks for MOS and 18.2 ks for pn). 
We processed the \XMM\ data using the Extended Source Analysis Software (ESAS), as integrated into the \XMM\ Science Analysis System (SAS; version 17.0.0), following the procedures in \cite{2019MNRAS.484.1946G}.
We reduced the raw event files from MOS and pn CCDs using tasks {\tt emchain} and {\tt epchain}, respectively. The solar soft proton flares were filtered out with {\tt mos-filter} and {\tt pn-filter}. The point sources were detected by task {\tt cheese} and then visually inspected and properly excluded. We used {\tt mos-spectra} and {\tt pn-spectra} to produce event images and exposure maps, as well as to extract spectra and response files. The instrumental background images and spectra were modeled with  {\tt mos$\_$back} and {\tt pn$\_$back}. We combined the event images, background images, and exposure maps from MOS and pn with {\tt comb}. We used {\tt adapt} to produce the final background subtracted, exposure corrected, and smoothed image (Fig.~\ref{fig:img}). 
The spectra from MOS/pn were fitted jointly with the {\tt XSPEC} package.
The nearby local background was used for fitting spectra.
We used the AtomDB (version 3.0.8) database of atomic data and the solar abundance table from \cite{2009ARA&A..47..481A}. The Galactic column density $N_{\rm H}=1.91\times10^{20}\ {\rm cm}^{-2}$ was from the NHtot tool \citep{2013MNRAS.431..394W}.

\subsection{\muse\ data processing}
The OC was observed with the Multi-Unit Spectroscopic Explorer (\muse; \citealt{2010SPIE.7735E..08B}) 
on the Unit Telescope 4 (Yepun) of the Very Large Telescope (VLT) during the nights of 2020 February 25 and 2020 March 17, 
under the European Southern Observatory (ESO) program 0104.A-0268(A) (PI: M. Sun). 
Both nights were clear for photometry with seeing of 0.$^{\prime\prime}$71 to 1.$''$38 (a median value of 0.$''$88).
Adopting the wide-field mode, four exposures (820 s each), in two slightly dithered positions, were taken for a total time of 0.91 h. The wavelength coverage is 4750 - 9350 \AA\ with a spectral resolution of $\sim$ 2600 at the wavelength of the OC's H$\alpha$ line.
We also carried out a sky background observation for 2 min at $\sim150''$ from the OC. 

The raw data of each pointing were reduced using the \muse\ pipeline (version 2.8.1; \citealt{2012SPIE.8451E..0BW,2020A&A...641A..28W}) 
with the ESO Recipe Execution Tool (EsoRex;  \citealt{2015ascl.soft04003E}), which performed the 
standard steps to calibrate the individual exposures and combine them into a datacube.
We also used the Zurich Atmosphere Purge software 
(ZAP; \citealt{2016MNRAS.458.3210S}) to improve the sky subtraction.
The CubeMosaic class implemented in the \muse\ Python Data Analysis 
Framework (MPDAF) package (\citealt{2016ascl.soft11003B}) was used to combine the individual datacube of each pointing into a final datacubes mosaic. 
Astrometry is calibrated with bright {\em 2MASS} stars in the field.

We used the public IDL software Kubeviz (\citealt{2016MNRAS.455.2028F})
to perform the spectral analysis for the final datacube mosaic. We first corrected the Galactic extinction by using the color excess 
from the recalibration (\citealt{2011ApJ...737..103S}) of the dust map of \cite{1998ApJ...500..525S}, adopting a Galaxy extinction law from 
\cite{1999PASP..111...63F} with $R_{V}=3.1$. 
Given the seeing value and the faintness of the diffuse emission,
we also smoothed the datacube with a Gaussian kernel of 6 pixels (or 1.$''$2). 
We fitted the H$\beta$, \OIII, \OI, \NII, H$\alpha$, 
and \SII\ emission lines with Gaussian profiles to obtain the emission-line fluxes, the velocity, 
and the velocity dispersion of the ionized gas. The two-dimensional maps of H$\alpha$ surface brightness, \NII/H$\alpha$ flux ratio, H$\alpha$ velocity, and velocity dispersion are shown in Fig.~\ref{fig:muse_maps}. We mask the spaxels with S/N $<$ 5 or velocity error and velocity dispersion error $>$ 50 \km.
We can also compare the total H$\alpha$ flux of the OC from the {\em MUSE} data with the result from \cite{2017ApJ...839...65Y} based on the narrow-band imaging data. We derive a total H$\alpha$ flux of (4.6$\pm0.1) \times 10^{-15}$ erg s$^{-1}$ cm$^{-2}$ from the full {\em MUSE} field. With the updated velocity and the \NII/H$\alpha$ flux ratio, we revise the total H$\alpha$ flux of \cite{2017ApJ...839...65Y} to (4.1$\pm0.2) \times 10^{-15}$ erg s$^{-1}$ cm$^{-2}$ for Orphan 1 and (4.8$\pm0.2) \times 10^{-16}$ erg s$^{-1}$ cm$^{-2}$ for Orphan 2.
The {\em MUSE} field of view (FOV) covers the whole Orphan 1 and part of Orphan 2, and the {\em MUSE} data are slightly deeper than the narrow-band imaging data, the {\em MUSE} H$\alpha$ flux is consistent with the H$\alpha$ flux from the {\em Subaru} narrow-band data.

\subsection{{\em APO/DIS} data processing}
We observed the OC with the Dual Imaging Spectrograph ({\em DIS}) on the Apache Point Observatory ({\em APO}) on 2020 January 30, 2021 January 16, and 2021 February 4 (PIs: C. Sarazin \& M. Sun). The first two nights were not photometric and a 6$'$ long slit with a width of 2$''$ was used.
The third night on 2021 February 4 was nearly photometric and a 6$'$ long slit with a width of 5$''$ was used.
On 2020 January 30, we observed the main body of the OC on two slit positions (70 min and 60 min respectively). These data are superseded by the later {\em MUSE} data but allow us to verify the velocity consistency, $7201\pm28$ \km\ from {\em DIS} vs. $7222\pm35$ \km\ from {\em MUSE} (uncertainty mainly from the uncertain {\em DIS} slit position) for the first slit position with the stronger detection than that of the other position. On 2021 January 16 and February 4, we observed two slit positions to the southeast of the OC that is outside of the {\em MUSE} field, in 70 and 40 min respectively.
The results are presented in Section~\ref{sec:kinematics}. 
The dome flats were used. All {\em DIS} velocities are calibrated with both the arc lamp spectra and night sky lines. The heliocentric correction was also made for all measured velocities.

\subsection{{\em Subaru} data processing}
In Fig~\ref{fig:img}, we used $g$, $r$, and net H$\alpha$ images.
The H$\alpha$-on data were obtained on 2017 May 27 with the N-A-L671
narrow-band filter of Suprime-Cam as an integration of thirty-four
$\leq 5$-min exposures with a total integration time of 165 min
under a natural seeing size of 0.$''$7 - 0.$''$9. The data were reduced
as described in \cite{2017ApJ...839...65Y}. We used astrometry.net (\citealt{2012ascl.soft08001L}) to obtain an astrometric solution.
Broadband images were obtained with the Hyper Suprime-Cam (HSC) in $r$ and
$g$ bands on 2016 March 10 and 2017 March 27. The number of
exposures, total exposure time, and typical seeing size were 11, 28.5
min and 0.$''$7 in $r$ band respectively, and 23, 66.5 min and 1.$''$0 in $g$ band respectively. The data
were reduced with HSCPipe version 4.0.5 (\citealt{2018PASJ...70S...5B}). We took the median
of all the exposures. The $r$ band data were also used for off-band of
H$\alpha$.
The on and off images were aligned and resampled with respect to WCS using SWarp (\citealt{2002ASPC..281..228B}). The off image was then scaled and subtracted from the on image to obtain the net H$\alpha$ image.
The remaining artifacts and stellar halo residuals were manually masked.

\begin{table}
\caption{Properties of the OC}
\begin{center}
\begin{tabular}{lc}
\hline
\hline
RA  & 11:44:23.2 \\
Dec & +20:11:00.2\\
$z$ & 0.0241 \\
radius (kpc) & 30 \\
SFR (${\rm M}_{\odot}$ yr$^{-1}$)& $<$ $10^{-3}$ \\
$kT_X$ (keV) & $1.6\pm0.1$\\
X-ray Abundance (${\rm Z}_{\odot}$) &  $0.14\pm0.03$\\
$L_{\rm 0.5-2\ keV}\ (10^{41} {\rm\ ergs\ s}^{-1})$ & $1.3$ \\
$L_{\rm bol}\ (10^{41} {\rm\ ergs\ s}^{-1})$ & $3.1$ \\
$n_{\rm e}$ ($10^{-3}f_{\rm X}^{-1/2}$ cm$^{-3}$) $^{a}$ & $3.1$ \\
$M_{\rm X}$ ($10^{10} {\rm\ M}_\odot$) & $1.0 (f_{\rm X} / 1.0)^{1/2}$ \\
$M_{\rm H\alpha}$ ($10^{7} {\rm\ M}_\odot$) $^{b}$ & $8.0 (f_{\rm H\alpha} / 0.01)^{1/2}$ \\
\hline
\end{tabular}
\end{center}
\begin{tablenotes}
\item
Note: $^{a}$: The average electron density of the X-ray gas. $f_{\rm X}$ and $f_{\rm H\alpha}$ are the filling factor of the X-ray and H$\alpha$ gas respectively.
$^{b}$: For the H$\alpha$ mass, the Case B recombination is assumed and we approximate the H$\alpha$ OC main body as a sphere with a radius of 8 kpc. Despite all the uncertainties, the mass of the warm gas is expected to be much smaller than that of the hot gas. 
\end{tablenotes}
\label{t:pro}
\end{table}

\section{results}
In general, the OC is $\sim 800$ kpc in projection from the centre of A1367 with $r_{500}\approx 900$ kpc derived from an average $T_X=3.5$ keV \citep{2002ApJ...576..708S} and $r_{500}-T_X$  relation \citep{2009ApJ...693.1142S}.
It is not far from the major axis of the cluster (NW - SE). It is also located near a cluster merger shock front \citep{2019MNRAS.486L..36G}. If it is truly located in the post-shock region, shock compression could have enhanced the density and X-ray luminosity of the OC, aiding to its discovery.
In X-rays, the OC peaks around the main H$\alpha$ OC, but with an offset of $\sim$ 12 kpc. As shown in Fig.~\ref{fig:muse_maps}, there seems to be an anti-correlation between the X-ray peak and the H$\alpha$ emission, with the X-ray peak surrounded by H$\alpha$ filaments.
There is an extension to the north but the analysis there is complicated by a bright background active galactic nucleus (AGN; more detail in Section~\ref{sec:agn}). There is also an X-ray extension to the SE, just like the H$\alpha$ OC. Our recent deep {\em Subaru} H$\alpha$ image of the field reveals more H$\alpha$ emission scattered around the X-ray OC, suggesting that there is a complex of warm, ionized clouds around the main H$\alpha$ OC discussed in \cite{2017ApJ...839...65Y}.
The positional coincidence of the H$\alpha$ clumps and the X-ray OC justifies their association. As discussed in Section~\ref{sec:origin}, the X-ray OC also cannot be a background cluster. The properties of the OC are summarized in Table~\ref{t:pro}.

\subsection{X-ray properties of OC}
The X-ray OC is asymmetric around its X-ray peak. Its umbrella-like morphology resembles the shape of a simulated isolated cloud moving in the ICM (e.g. \citealt{2020MNRAS.499.5873C}). A radial surface brightness profile centred on its peak shows an effective radius of $\sim$ 30 kpc (Fig.~\ref{fig:sbp}). It has a lower temperature than that of the surrounding ICM (1.6 keV vs. 2.9 keV) from \XMM\ data. The best-fit abundance from the single-$T$ model is only $\sim$ 0.14 solar but that is biased low due to the intrinsically multi-$T$ gas in the OC. The total X-ray bolometric luminosity is $3.1\times10^{41} {\rm\ ergs\ s}^{-1}$, comparable to those of massive cluster galaxies (e.g. \citealt{2007ApJ...657..197S}).
The cooling time of the X-ray gas in the OC is more than 3.6 Gyr so the warm gas is not the product of cooling in the soft X-ray gas. Instead, the X-ray OC likely glows because of the mixing between the cold gas and the surrounding hot ICM. 
More details are presented below.

\subsubsection{Spectral Properties of the X-ray OC}
We extracted the spectra of the OC from the \XMM\ data, excluding the bright point source near the northern edge of the OC. We also extracted the spectra of the immediate surroundings as the local background.
We emphasize that the mixing between stripped cold ISM and hot ICM can produce the multi-phase gas as for the case of OC. However, physically motivated X-ray spectral models to study the mixing clouds are unavailable. Nevertheless, we can still gain insight
with simple models.
The spectra are fitted with different models in {\tt XSPEC}, with results detailed in Table~\ref{t:fit}.
The single-$T$ model gives a very low abundance, which is most likely the result of intrinsically multi-$T$ gas in the OC (e.g. \citealt{2010ApJ...708..946S}). Including an additional power-law model (for X-ray point sources unresolved by \XMM) or using a two-$T$ model results in better fits, because these models include more free parameters to provide a better approximation. We also tried a multi-temperature model ({\tt CEMEKL}), first used on stripped tails by \cite{2010ApJ...708..946S}. The maximum temperature of {\tt CEMEKL} is fixed to that of the surrounding ICM, and its abundance is fixed to the typical value of the ICM (0.3 solar). The better fitting statistic from the {\tt CEMEKL} model also suggests a multi-$T$ nature of the OC. However, all these models only provide over-simplified and phenomenological fits to the X-ray OC, given the limited angular resolution and the limited statistics of the \XMM\ data.
On the other hand, the best-fit temperature from the one-$T$ model can be taken as the spectroscopic or effective temperature of the X-ray OC, and can be compared with the temperature of other multi-phase gas like the stripped tails. The X-ray luminosity from these models is robust as e.g., {\tt APEC} and {\tt CEMEKL} models give consistent X-ray luminosity.
 
\begin{table*}
\caption{X-ray spectral models for the OC.}
\begin{center}
\begin{tabular}{lcc}
\hline
\hline
Model & Parameters & C-stat/d.o.f \\
\hline
{\tt APEC}  & $kT=1.6\pm0.1$, $Z=0.14\pm0.03$, $L_{T}=1.3\pm0.1$ & 269/219 \\
{\tt APEC+PL} & $kT=1.1\pm0.1$, $Z=0.21\pm0.03$, $L_{T}=7.4\pm0.4$, $\Gamma=(1.7)$, $L_{P}=1.5\pm0.1$ & 231/216 \\
{\tt APEC+APEC} & $kT_1=1.0\pm0.1$, $Z_1=(0.3)$, $kT_2=3.6\pm0.7$, $Z_2=(0.3)$, $N_1/N_2=0.36$ & 244.7/216 \\
{\tt APEC+APEC} & $kT_1=0.97\pm0.04$, $Z_1=(1.0)$, $kT_2=2.7\pm0.3$, $Z_2=(0.3)$, $N_1/N_2=0.1$ & 250.6/216 \\
{\tt CEMEKL} & $\alpha=1.5\pm0.2$, $kT_{\rm max}=(2.9)$, $Z=(0.3)$, $L_{T}=1.3\pm0.1$ & 240.3/220 \\
\hline
\end{tabular}
\end{center}
\begin{tablenotes}
\item
{\sl Note:}
The Galactic absorption ($1.91 \times 10^{20}\ {\rm cm}^{-2}$) is included in all cases with a model of {\tt TBABS}.
$L_T$ (0.5-2 keV) and $L_P$ (2-10 keV) are the luminosity of {\tt APEC}/{\tt CEMEKL} and power-law model with unit of $10^{41}$ erg s$^{-1}$. The unit for $kT$ is keV and the unit for the abundance $Z$ is solar. Parameters in parentheses are fixed.
{\tt CEMEKL} is a multi-temperature plasma emission model with emission measures following a power-law distribution in temperature: ${\rm EM}(T) \propto (T/T_{\rm max})^\alpha$.
\end{tablenotes}
\label{t:fit}
\end{table*}

\subsubsection{Bright X-ray point source in the X-ray OC}\label{sec:agn}
From the \XMM\ spectra, the bright X-ray point source near the northern edge of the OC is best fitted with a power-law model, with a photon index of $\Gamma=1.9\pm0.1$ and a flux of $f_{2-10{\rm\ keV}}=5.9\times10^{-14}{\rm\ ergs\ cm^{-2}\ s^{-1}}$.
Its faint optical counterpart (SDSS~J114425.15+201219.6) was selected as an AGN candidate \citep{2015ApJS..219...39R}.
The ${\rm log}N-{\rm log}S$ relation \citep{2008A&A...492...51M} predicts an X-ray source density of 20 deg$^{-2}$ above the flux of this source.
Indeed, this X-ray point source is the brightest one within a radius of 8$'$ in the \XMM\ FOV, which corresponds to 18 deg$^{-2}$. Thus, this source is most likely a background AGN unrelated to the OC.

\subsubsection{Gas density and mass of the X-ray OC}
\begin{figure}
\begin{center}
\includegraphics[angle=0,width=0.49\textwidth]{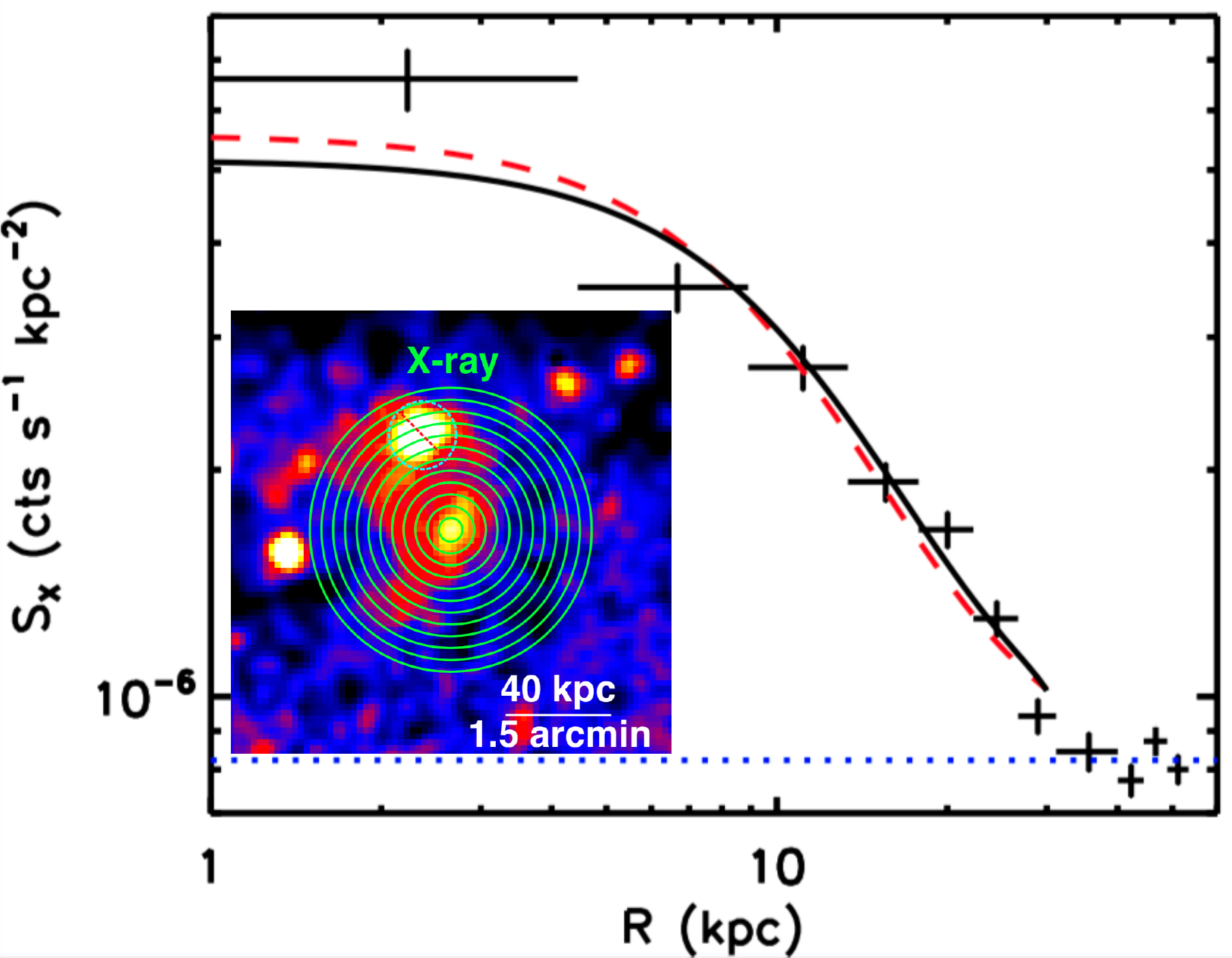}
\caption{
The surface brightness profile of the OC. The solid black line is the best fit of a $\beta$-model convolved with the \XMM\ PSF, while the dashed red line is the $\beta$-model without correction for the PSF. The dotted blue line is the local background level. 
The inset in the left corner shows the annuli for the SBP extraction. The annuli are centred on the X-ray peak.
}
\label{fig:sbp}
\end{center}
\end{figure}

We estimate the hot gas density of the OC from the {\tt XSPEC} normalization, assuming a spherical cloud of uniform density.
The {\tt apec} normalization $\eta$ is
\begin{equation}
\eta=\frac{10^{-14}}{4\pi[D_{\rm A}(1+z)]^2}\int n_{\rm e}n_{\rm H}dV
\end{equation}
where $z=0.022$ is the redshift of A1367, $D_{\rm A}$ is the angular size distance at $z=0.022$,
$n_{\rm e}$ and $n_{\rm H}$ are electron and proton densities. Because the X-ray shape of the OC is asymmetric, we use an effective radius of $R_{\rm OC} \sim 30$ kpc enclosing most of its diffuse X-ray emission. Fig.~\ref{fig:sbp}  shows the radial surface brightness profile (SBP) of the OC centred at its X-ray peak.
The SBP also suggests that the diffuse X-ray emission extends to around 30 kpc.
The resultant average density is $n_{\rm e}=3.1\times 10^{-3}f^{-1/2}{\rm\ cm^{-3}}$, where
$f$ is the filling factor of the X-ray emitting gas.
The gas mass of the OC is $M_{\rm OC}=1.0\times10^{10}f^{1/2}{\rm\ M}_\odot$ 
for uniform density.
We also try a $\beta$-model \citep{1976A&A....49..137C} convolved with \XMM\ PSF to fit the SBP of OC as in Fig.~\ref{fig:sbp}.
The $\beta$-model gas distribution is given by $n_{\rm gas}(r)=n_0[1+(r/r_c)^2]^{-3\beta/2}$, which is an analytical model with the derived X-ray SBP also following  a $\beta$-model in the form of $I_{\rm X}(r)=I_0[1+(r/r_c)^2]^{1/2-3\beta}$.
We use the analytical formula Eq.~(10) of \cite{2016MNRAS.459..366G} to convert the central surface brightness $I_0$ (from the $\beta$-model fitting to the SBP) to the central gas density $n_0$.
The related central electron density is $4.6\times 10^{-3}f^{-1/2}{\rm\ cm^{-3}}$.
The gas cooling time $t_{\rm cool}$ at $n_{\rm e} = 4.6\times 10^{-3} {\rm\ cm^{-3}}$ and $kT$ = 1.6 keV and $Z$ = 0.5 is $t_{\rm cool}= 3.6$ Gyr.
The gas cooling time at $n_{\rm e} = 3.1\times 10^{-3} {\rm\ cm^{-3}}$ and $kT$ = 1.6 keV and $Z$ = 0.14 is $t_{\rm cool}=7.2$ Gyr.

\subsection{H$\alpha$ properties of OC}
The H$\alpha$ OC is composed of the main body covered by our {\em MUSE} observations, a SE trail, and some other clumps around the X-ray OC (Fig.~\ref{fig:img}). Our new {\em MUSE} observations not only confirm the association of the H$\alpha$ OC with A1367, but also provide details on the kinematics and line diagnostics of the cloud, as shown in Fig.~\ref{fig:muse_maps}. There is a clear velocity gradient in the main body of the H$\alpha$ OC and the velocity dispersion is typically small, $\sim$ 80 \km. Line diagnostics (see detail in Section~\ref{sec:line}) suggest little SF in the OC but the ionization mechanism remains unclear.
Why is the brightest H$\alpha$ emission offset from the brightest X-ray emission?
OC is likely in a late evolutionary stage of mixing between the stripped cold ISM and the hot ICM as suggested below.
The bright H$\alpha$ clumps may be associated with the only surviving cold clouds while the bulk of the X-ray OC is free of cold gas now. Future \hi\ and CO observations of the OC will be important to understand the evolution of the OC.

\begin{figure*}
\begin{center}
\includegraphics[angle=0,width=1\textwidth]{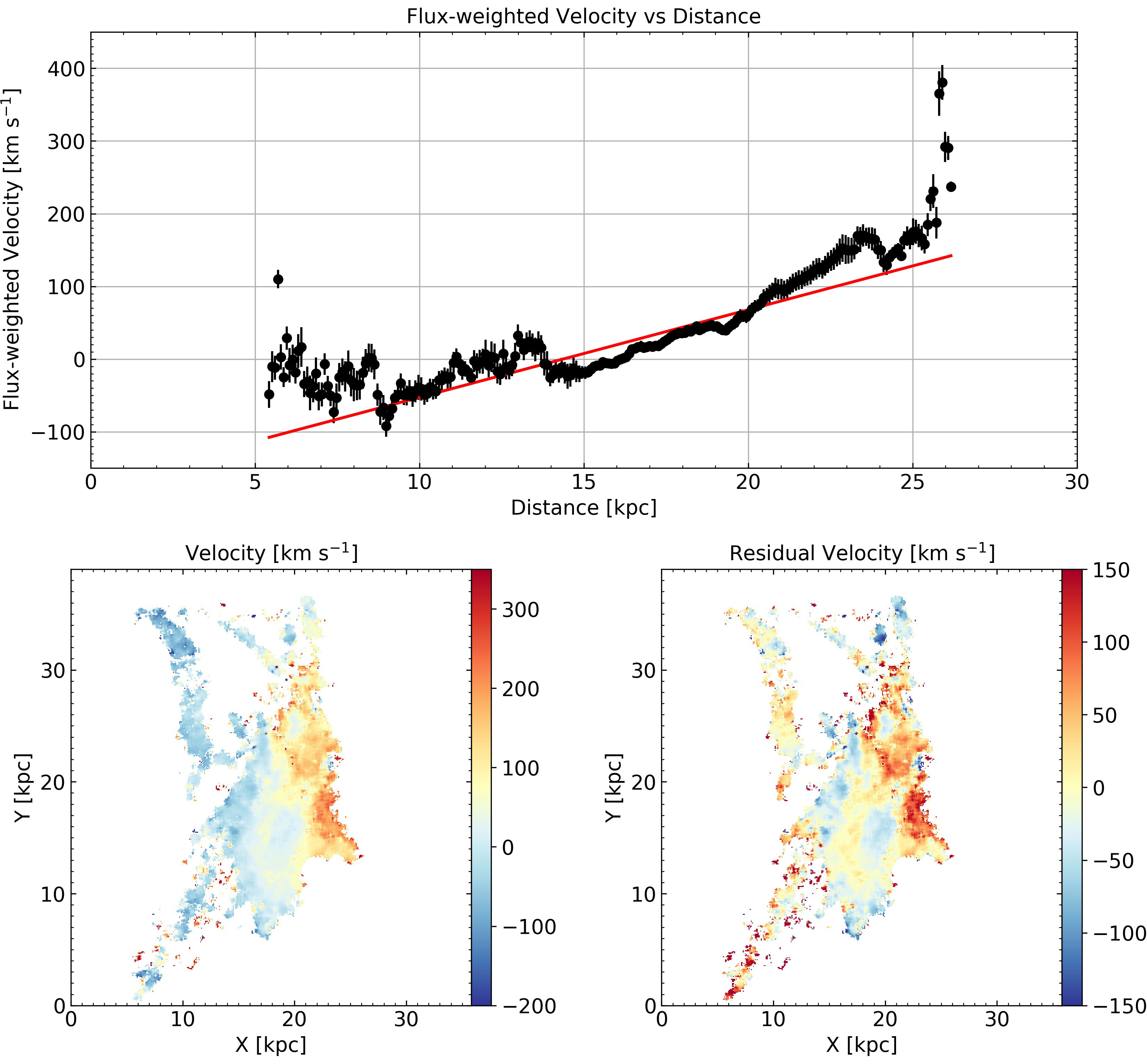}
\caption{The best-fit results of the velocity gradient for the ionized gas in the {\em MUSE} field.
During the fitting process, we rotated the velocity map at a certain angle and calculated the flux-weighted average of the velocity in each column. The best-fit direction of the velocity gradient is determined to have the lowest root mean square (RMS) in the corresponding residual velocity map (velocity map - velocity gradient). 
Such an analysis determined the best-fit direction of the velocity gradient, 7.1 deg clockwise from the west as shown in Fig.~\ref{fig:muse_maps}. The {\em MUSE} velocity map is then rotated accordingly to align the velocity gradient east-west here.
\textit{\textbf{Upper panel}}: the H$\alpha$ flux weighted velocity as a function of the distance from the eastern edge of the {\em MUSE} field, after the rotation. The black dots show the averaged velocities, weighted by the H$\alpha$ flux from spaxels in each column. 
The red solid line shows the best-fit velocity gradient. 
\textit{\textbf{Lower panels}}: the rotated velocity map and the corresponding residual map after removing the best-fit velocity gradient. 
}
\label{fig:vel_gradient}
\end{center}
\end{figure*}

\subsubsection{Kinematics of the H$\alpha$ OC}\label{sec:kinematics}
From the integrated spectrum of the whole OC in the {\em MUSE} field, the redshift is measured to be 0.0241.
The OC is likely moving westward as suggested by the umbrella-like X-ray morphology and the SE H$\alpha$ trail (Fig.~\ref{fig:img}; \citealt{2017ApJ...839...65Y}).
The H$\alpha$ velocity map in Fig.~\ref{fig:muse_maps} shows a nearly east-west velocity gradient.
We estimate the velocity gradient by minimizing the velocity residuals relative to a model with a constant gradient from the {\em MUSE} velocity map.
An angle of 7.1$\pm$1.0 deg clockwise from the west, as shown in Fig.~\ref{fig:muse_maps}, results in the minimal velocity residual. The velocity gradient along this direction is substantial, 12 \km\ per kpc.
The H$\alpha$ OC has a total velocity gradient of $\sim$ 200 \km\ nearly aligned east-west (Fig.~\ref{fig:muse_maps} and Fig.~\ref{fig:vel_gradient}). Such a large velocity gradient is higher than those typically found in isolated \hi\ clouds (e.g. \citealt{2015AJ....149...72C}).
We can also estimate the cloud’s dynamical mass if we assume that the velocity gradient is due to rotation in a stripped disk. We extrapolate the velocity gradient (12 \km\ per kpc) to the OC's radius of 30 kpc, then the dynamic mass is $M=v^2r/G=9.0 \times 10^{11}\ {\rm M}_\odot$. Thus, if the observed velocity gradient of the OC is the imprint of the rotation in the disk of its parent, its parent must be a massive galaxy. On the other hand, the rotation pattern in the stripped ISM is not expected to be conserved for a long period of time after stripping (e.g. \citealt{2021A&A...646A.139B}).

While the map of the velocity dispersion is shown in Fig.~\ref{fig:muse_maps}, we also
spatially divided the {\em MUSE} FOV into nine large regions (see the upper right panel of Fig.~\ref{fig:muse_maps}) and examined the velocity dispersion there. The velocity dispersion in these regions ranges from 50 to 149 \km, with a median value of $\sim$ 80 \km.
There are several positions to the west side of the OC with velocity dispersion as high as $\sim$ 180 \km\ but the typical velocity dispersion is small.
We can also derive the velocity dispersion of the cloud at $\sim$ 10 kpc scales from the velocity map shown in Fig.~\ref{fig:muse_maps}.
The standard deviations of the velocity histogram, weighted or not weighted by the H$\alpha$ flux, are 75 \km\ and 88 \km, respectively. If the average velocity gradient of the OC is subtracted, those values decrease to 55 \km\ and 67 \km, respectively. The above analysis examines the velocity dispersion of the warm gas at kpc - 10 kpc scales, indicating the small contribution from turbulence at those scales, at least in the warm gas. 

\begin{figure}
\begin{center}
\includegraphics[angle=0,width=0.46\textwidth]{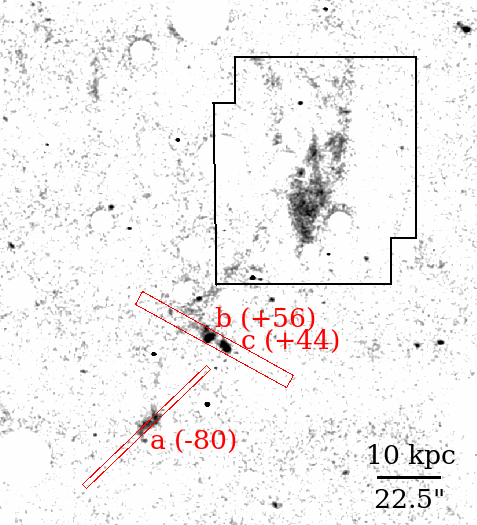}
\caption{The {\em APO/DIS} slit positions at the southeast of the H$\alpha$ OC. The velocities (in a unit of \km, relative to $z=0.024$ as in Fig.~\ref{fig:muse_maps}) are also shown. The image shows the net H$\alpha$ emission from {\em Subaru}, with stars and galaxies masked out. The {\em MUSE} FOV is also shown in a black solid box.
}
\label{fig:dis_slits}
\end{center}
\end{figure}

We also obtained a few more velocities for the warm, ionized gas beyond the {\em MUSE} field from {\em APO/DIS}, as shown in Fig.~\ref{fig:dis_slits}.
The three positions and the measured velocities are listed:
a at (11:44:26.8 +20:09:16.3) -- $7115\pm37$ \km, b at (11:44:25.3 +20:09:46.3) -- $7239\pm29$ \km, c at (11:44:24.9 +20:09:43.1) -- $7251\pm26$ \km. Regions b and c are most likely \hii\ regions from their high surface brightness (easily detected in 10 min with {\em DIS}), as also suggested by \cite{2017ApJ...839...65Y}. \NII\ and \SII\ lines are also detected in regions b and c. We constrained \NII/H$\alpha$ $\sim$ 0.4 and \OI/H$\alpha$ $<$ 0.2.
As shown in Fig.~\ref{fig:img}, there are more H$\alpha$ clumps around the OC. 2MASX~J11443212+2006238 with an 85 kpc H$\alpha$ is also nearby and has a similar velocity of 7214 \km\ \citep{2017ApJ...839...65Y,2017A&A...606A.131G}.

\subsubsection{Line diagnostics of the H$\alpha$ OC}\label{sec:line}
\begin{figure*}
\begin{center}
\includegraphics[angle=0,width=1\textwidth]{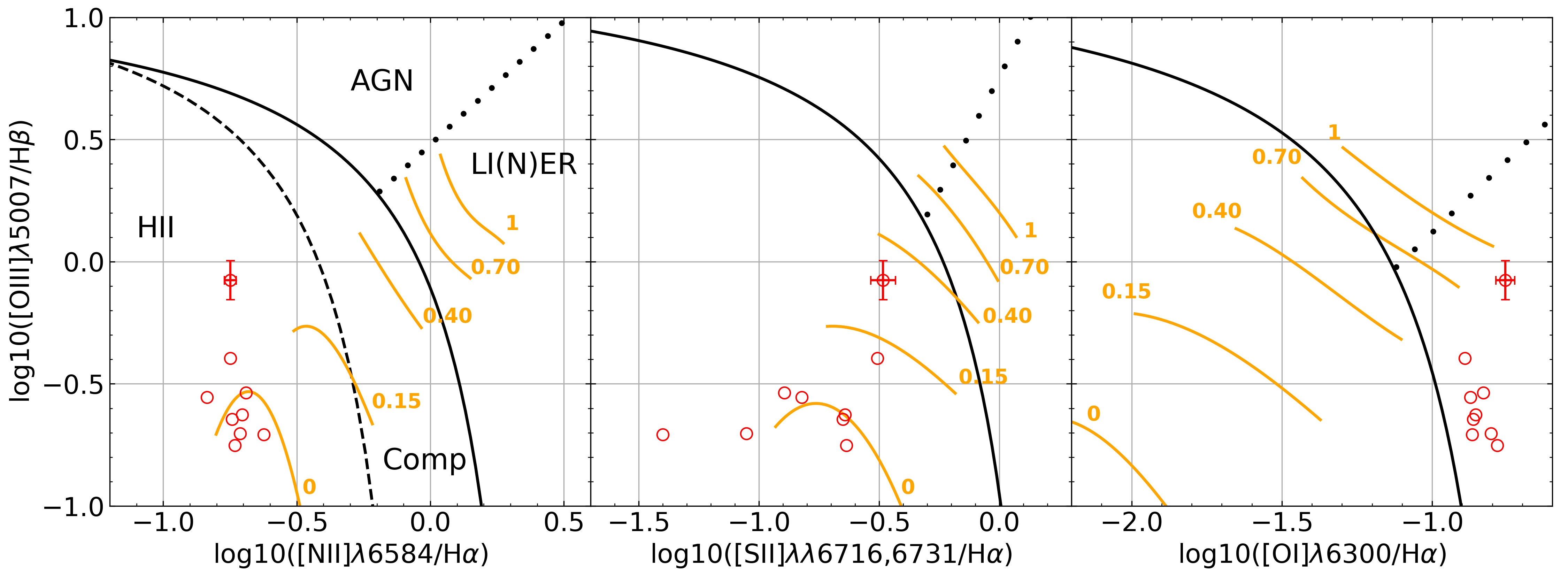}
\caption{The emission line ratios in 9 large spatial regions of the A1367 OC.
The dashed, solid and dotted lines show the criteria (from \protect\citealt{2003MNRAS.346.1055K}, \protect\citealt{2001ApJ...556..121K} and \protect\citealt{2010MNRAS.403.1036C}) to separate the regions
of SF, composite (Comp), AGN, and LI(N)ER, respectively. 
Typical errors in the ratios are plotted on 
the data points with the largest \OIII/H$\beta$ ratio (from the region at $\Delta$RA $\sim$ 10$''$ and $\Delta$DEC $\sim$ 8$''$ in Fig.~\ref{fig:muse_maps}).
The solid orange lines show the fractions 
(from 0 to 1) of H$\alpha$ flux from radiative shocks as predicted by the models in \protect\cite{2011ApJ...734...87R}.
While the first two plots may suggest these regions as \hii\ regions, the \OI/H$\alpha$ ratios in these regions are too high and there is no evidence of SF from the {\em GALEX} data.
}
\label{fig:muse_bpt}
\end{center}
\end{figure*}

We used emission-line diagnostics to examine the excitation mechanisms for the warm ionized gas in the OC. In order to enhance the S/N of faint emission lines, we again focus on those nine large regions shown in Fig.~\ref{fig:muse_maps}
and measure the H$\beta$, \OIII, \OI, \NII, H$\alpha$, and \SII\ emission-line fluxes from the co-added spectra within each region. 
The corresponding emission-line flux ratios are shown in Fig.~\ref{fig:muse_bpt}.
We used the criteria from \cite{2001ApJ...556..121K,2003MNRAS.346.1055K}
to classify the AGN, composite, and star-forming regions. The demarcation of \cite{2010MNRAS.403.1036C} was used to separate the Seyfert and low-ionization (nuclear) emission-line regions (LI(N)ERs). As shown in Fig.~\ref{fig:muse_bpt}, 
while the \NII/H$\alpha$ and \SII/H$\alpha$ flux ratios are low, the \OI/H$\alpha$ flux ratios are relatively high and the \OIII/H$\beta$ 
flux ratios are also low, indicating the LI(N)ER-like emission of the ionized gas. This is a good example of LI(N)ER-like emission not in an active nucleus, but in an extragalactic region (also see \citealt{2012ApJ...749...43Y,2017A&A...606A..83C} for similar examples in stripped tails still close to the parent galaxy).

\subsection{H$\alpha$ - X-ray correlation for the OC}
\begin{figure*}
\begin{center}
\includegraphics[angle=0,width=0.98\textwidth]{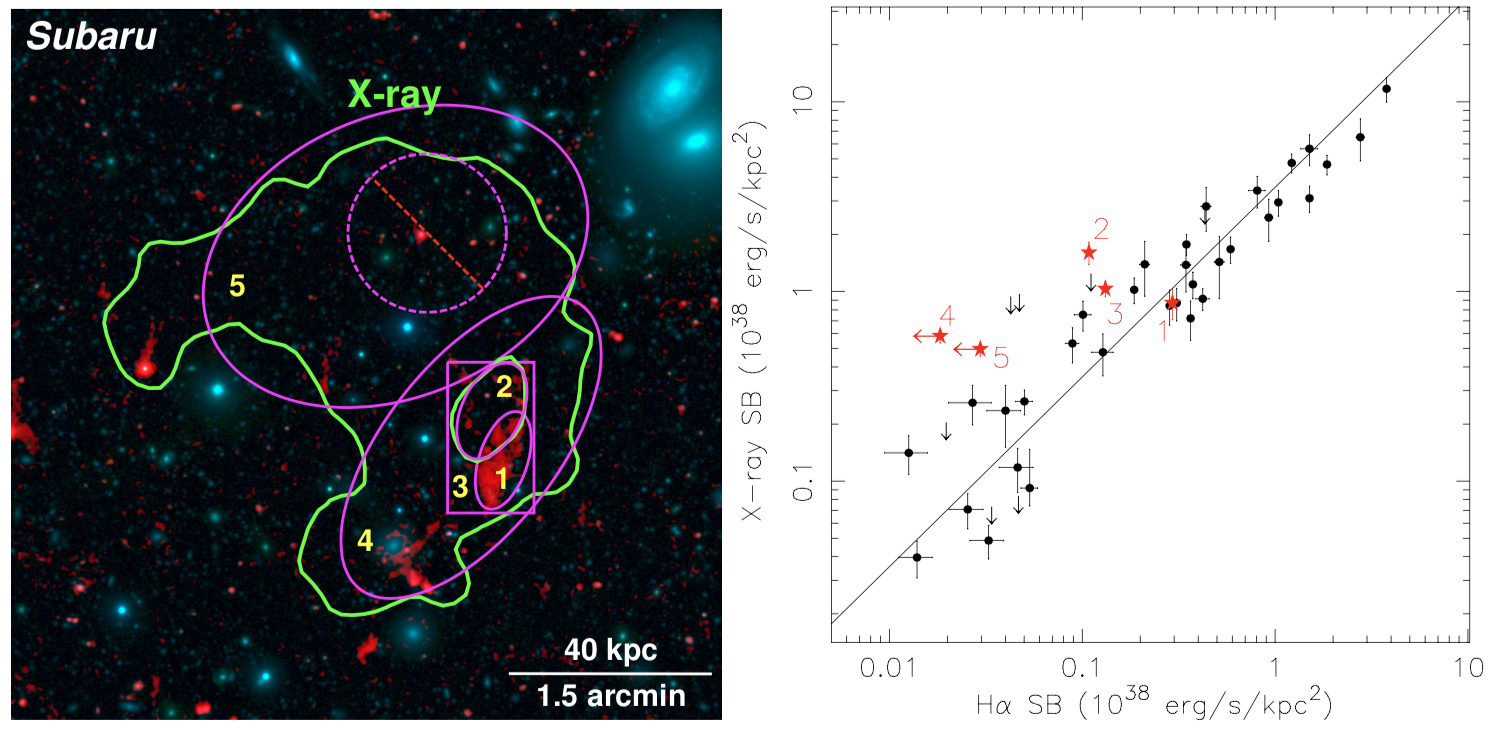}
\caption{\textit{\textbf{Left panel}}:
regions used to study the H$\alpha$ - X-ray correlation. Regions 1 and 2 are ellipses around the H$\alpha$ and X-ray peaks, respectively. Region 3 is a box excluding regions 1 and 2. Region 4 is an ellipse excluding region 3. Region 5 is an ellipse excluding the bright X-ray source and region 4. For the H$\alpha$ emission, galaxies, background sources, and \hii\ regions are all excluded. The same {\em Subaru} three-color composite image as shown in Fig.~\ref{fig:img} is shown here, with the X-ray contours in green.
\textit{\textbf{Right panel}}: the measured H$\alpha$ and X-ray surface brightnesses for these five regions in red are plotted with the best-fit linear relation and all data points in black from \protect\cite{2021arXiv210309205S}. For the A1367 OC, while the correlation around the H$\alpha$ cloud is consistent with those for stripped tails still attached to their parent galaxies, the X-ray-to-H$\alpha$ ratios are typically higher, which should not be a surprise for the OC in a much more advanced stage of evolution than the stripped tails. Note that only generous upper limits on the total H$\alpha$ emission are put for regions 4 and 5.
}
\label{fig:regions}
\end{center}
\end{figure*}

We also examined the diffuse H$\alpha$ - X-ray correlation for the OC to compare with the tight correlation recently found for stripped tails still attached to their parent galaxies \citep{2021arXiv210309205S}. Such a tight correlation supports the mixing of the stripped ISM with the hot ICM as the origin of the multi-phase stripped tails.
Five regions are selected (Fig.~\ref{fig:regions}). H$\alpha$ and X-ray surface brightnesses are measured in these regions, with emission from galaxies, background sources, and \hii\ regions excluded. The bolometric X-ray flux in individual regions is from the spectral fitting with nearby local background (mostly from the ICM emission) subtracted. The H$\alpha$ emission in regions 1 - 3 is robustly measured. There is some diffuse H$\alpha$ emission in region 4, e.g. the diffuse tail, but \hii\ regions (b and c in Fig.~\ref{fig:dis_slits}) and galaxies are removed. Some faint, diffuse H$\alpha$ emission may also be present in region 5.
However, the level of faint, diffuse H$\alpha$ emission beyond the main body is quite uncertain, because of the uncertainty of the flat fielding at large scales and the subtraction of the light from other objects.
Thus, only upper limits are estimated for regions 4 and 5. As shown in Fig.~\ref{fig:regions}, away from the H$\alpha$ OC, the X-ray-to-H$\alpha$ ratio is elevated, as expected for a cloud that has long left its parent galaxy and evolved in the ICM for a long time.
Most cold gas are already gone so active mixing may only proceed around the H$\alpha$ OC.
Is the high X-ray/H$\alpha$ ratio related to the weak SF activity in the OC? Many stripped tails in the \cite{2021arXiv210309205S} sample have very weak SF comparable to that in the OC (e.g. NGC~4569, ESO~137-002, CGCG 097-073, CGCG 097-079, and D100), but the X-ray/H$\alpha$ ratios in their tails are all similar to the median value from \cite{2021arXiv210309205S}. Moreover, SF in the OC is outside of the main H$\alpha$ OC but the X-ray/H$\alpha$ ratio in the main H$\alpha$ OC is the lowest among all regions of the OC. Thus, the weak SF in the OC should not account for its large X-ray-to-H$\alpha$ ratio.

The mean temperature of the whole X-ray OC, $1.6\pm0.1$ keV, is higher than typical temperatures of X-ray tails of cluster late-type galaxies ($\sim0.9$ keV, \citealt{2021arXiv210309205S}), which may also suggest an advanced evolutionary stage of the X-ray OC as it mixes with the surrounding hotter ICM.
We note that the elevation of X-ray emission and temperature of OC might be caused by the merger shock if the OC is truly in the post-shock region. The shock Mach number is $M\sim1.6$ \citep{2019MNRAS.486L..36G}, which can produce a temperature jump of $T_J=1.6$ (i.e. from 1.0 keV to 1.6 keV).

\section{Discussion}
\subsection{Origin of the OC}\label{sec:origin}
Firstly, we check if the X-ray OC is a background galaxy cluster.
We examined the X-ray spectral properties of the OC assuming different $z$, exceeding 0.024. For each assumed $z$, the best-fit $T_X$ and $L_X$ are derived. 
If the OC is a background cluster, it should lie on the $L_X - T_X$ relation for groups and clusters (e.g. \citealt{2016A&A...592A...3G}). This analysis
constrains the redshift to the range $0.14 < z < 0.29$ for the X-ray OC (1$'$ = 148 - 261 kpc at this $z$ range).
$L_*$ galaxies in clusters in this redshift range should have an $r$-band magnitude of 17.8 - 19.6 AB mag, well within the detection limit of {\em SDSS}. However, within 150 kpc of the X-ray peak (for $z = 0.14 - 0.29$), none of the {\em SDSS} sources with a `type' of GALAXY is brighter than the above $r$-band magnitude. This is also shown from our deep {\em Subaru} data at the $B$, $R$ (including HSC's $r $-band) and $i$ bands. Since the brightest central galaxy (BCG) is typically more than 5 times more luminous than $L_*$, we can rule out the scenario that the X-ray OC is a background cluster.

Then the origin of the OC may be an infalling galaxy group or the stripped ISM from a massive galaxy.
For the temperature of the OC ($1.6$ keV), the expected X-ray luminosity is $L_{\rm bol}=7.1\times10^{42}$ erg s$^{-1}$ from the $L_X-T_X$ relation of galaxy clusters and groups (e.g. \citealt{2016A&A...592A...3G}), which is over 20 times higher than the observed value.
Can the OC be the remnant of an infalling galaxy group? A galaxy group this massive almost always has a BCG more luminous than $L_*$.
However, we examined the {\em 2MASS} Extended Source Catalog and found no E/S0 galaxies of brighter than 0.8 $L_*$ \citep{2001ApJ...560..566K} within 0.5 $r_{\rm 500, infall group} \approx 300$ kpc of the OC (from $r_{500}-T_X$ relation; \citealt{2009ApJ...693.1142S}).
Moreover, the remnant X-ray core of an infalling galaxy group typically does not have associated H$\alpha$ emission. Thus, it is unlikely the OC is a remnant core of an infalling galaxy group.  
It's more likely that the OC originates from the stripped ISM of an infalling galaxy.
The parent galaxy should not be small,
giving the significance of the X-ray gas mass of the OC ($\sim 10^{10}\ {\rm M}_\odot$).
It may not be accidental to find the OC in the NW of A1367, because A1367 is located in a node of the cosmic web.
Several galaxy groups with a higher fraction of SF galaxies are falling into it, especially in the NW direction, and the stripping processes may be very active there \citep{2004A&A...425..429C}.
Galaxies with stripped \hi\ gas \citep{2018MNRAS.475.4648S} and H$\alpha$ tails \citep{2017ApJ...839...65Y} preferentially gather around in the same region of the cluster.

Isolated H$\alpha$ clouds like the OC are rare in galaxy clusters, e.g. none found in the H$\alpha$ surveys in the Coma cluster, A851, and CL0024+17 \citep{2010AJ....140.1814Y,2015AJ....149...36Y}.
The only other isolated H$\alpha$ cloud in a galaxy cluster we are aware of is SECCO~1 in the Virgo cluster \citep{2017MNRAS.465.2189B,2017ApJ...843..134S,2018MNRAS.476.4565B}. 
SECCO~1 is a faint, star-forming stellar system with some diffuse H$\alpha$ emission.
Its physical size, $\sim$ 1.2 kpc in radius for each of the two pieces \citep{2018MNRAS.476.4565B}, is much smaller than the OC discussed in this paper. It has a rather high metallicity of $\sim$ half solar for its low optical luminosity
\citep{2017MNRAS.465.2189B,2017ApJ...843..134S}. \cite{2017MNRAS.465.2189B} suggested that SECCO~1 was formed from a pre-enriched gas cloud, possibly stripped from a massive galaxy in the Virgo cluster. There is no report of an X-ray counterpart of SECCO~1 and the properties of SECCO~1 appear very different from those of the A1367 OC.
Future wide-field H$\alpha$ surveys (e.g. VESTIGE, \citealt{2018A&A...614A..56B}) should be able to constrain the abundance of isolated H$\alpha$ clouds in clusters.

\subsection{Pressure balance in the X-ray OC}
\begin{figure*}
\begin{center}
\includegraphics[angle=0,width=0.99\textwidth]{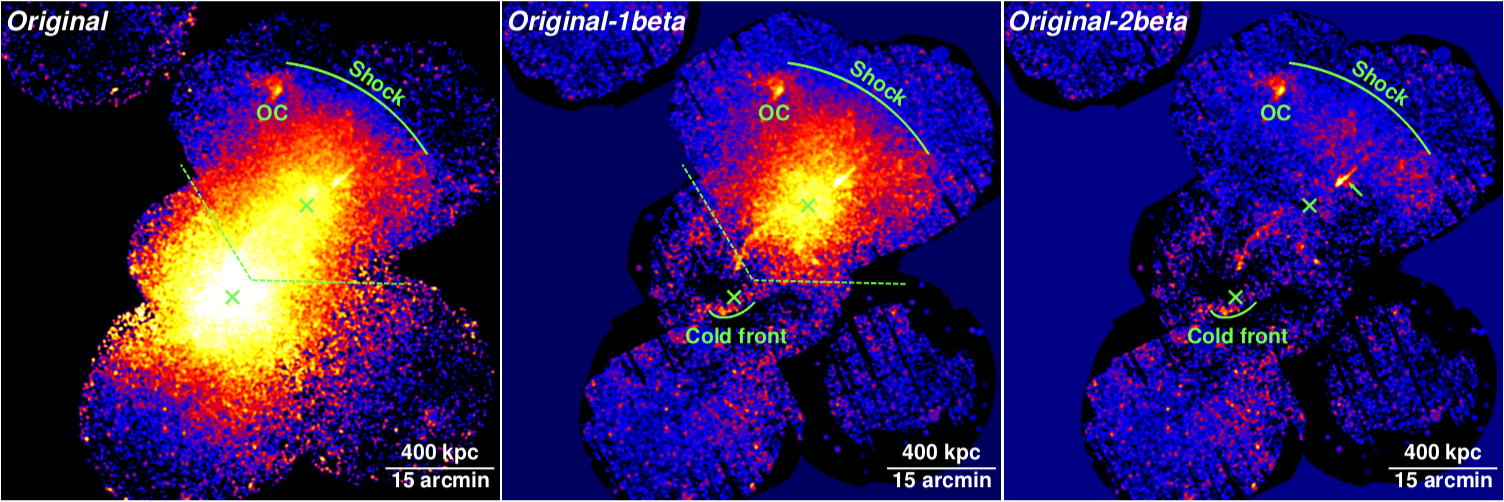}
\caption{
Original and residual X-ray images of A1367.
\textit{\textbf{Left panel:}} the 0.7 - 1.3 keV \XMM\ mosaic of A1367, with background subtracted and exposure corrected. X-ray point sources are also removed in this mosaic to show the diffuse emission better. The dashed line marks the boundary between the SE and NW subclusters, for the purpose of double $\beta$-model fit to the SBP of the cluster. Two green crosses mark the centres of two $\beta$-models for the SBP fitting. The positions of the OC and the shock front are also marked.  
\textit{\textbf{Middle panel:}} residual emission after subtracting the first $\beta$-model centred on the SE subcluster. Regions (e.g. CCD edges and gaps) with low exposure time and limited statistics have been removed. The NW subcluster and a cold front to the south of the SE subcluster core are significant. 
\textit{\textbf{Right panel:}} residual emission after further subtracting the second $\beta$-model centred on the NW subcluster. The long X-ray tail of UGC~6697 is marked by a green arrow. The OC also appears as a significant residual feature.
}
\label{fig:dif}
\end{center}
\end{figure*}

A1367 is undergoing a merger along the NW-SE direction (e.g. \citealt{2002ApJ...576..708S,2019MNRAS.486L..36G}). We can approximate its X-ray surface brightness distribution with two superimposed $\beta$-models of $I=I_{01}(1+r_1^2/r_{\rm c1}^2)^{1/2-3\beta_1}+I_{02}(1+r_2^2/r_{\rm c2}^2)^{1/2-3\beta_2}$, each centred on a subcluster as shown in Fig.~\ref{fig:dif}.
Before we fit the SBP of the SE subcluster with a $\beta$-model, we mask out the NW subcluster beyond the dashed line (0 deg to 120 deg counterclockwise from the west) in Fig.~\ref{fig:dif}.
Then the image of the NW subcluster is obtained by subtracting the first $\beta$-model from the original diffuse cluster image as shown in Fig.~\ref{fig:dif} middle panel.
We fit the second $\beta$-model to the NW subcluster. The Fig.~\ref{fig:dif} right panel shows the residual emission after subtraction these two $\beta$-models from the original image. While the residual large-scale features may be sensitive to the model properties (e.g. centroid, asymmetry), the small residual features are robust. The residual image of Fig.~\ref{fig:dif} right panel reveals some significant features, including a cold front in the SE subcluster \citep{2010A&A...516A..32G}, a long 
X-ray tail of UGC~6697 \citep{2021arXiv210309205S},
long X-ray trails associated with the Blue Infalling Group \citep{2017ApJ...839...65Y,2019MNRAS.484.2212F} and the X-ray emission of the OC.

We then derive the ICM density distribution from the best-fit $\beta$-model to the cluster SBPs.
We note that this method assumes the X-ray surface brightness is proportional to the emission measure as $EM_{\rm model}=\int (n_{\rm e1}^2+n_{\rm e2}^2)dl$, where $n_{\rm e1}=n_{\rm 01}(1+r_1^2/r_{\rm c1}^2)^{-3\beta_1/2}$ and $n_{\rm e2}=n_{\rm 02}(1+r_2^2/r_{\rm c2}^2)^{-3\beta_2/2}$ are the electron densities of SE and NW subclusters, while the true X-ray surface brightness, assuming two subclusters are merging on the plane of the sky, is proportional to $EM_{\rm true}=\int (n_{\rm e1}+n_{\rm e2})^2dl$. There is a discrepancy between $EM_{\rm model}$ and $EM_{\rm true}$, especially between the two peaks of subclusters. However, at the far sides from each peak, the SBP is dominated by the gas density of each subcluster. We can correct the density normalization through comparing the SBP from X-ray observations and the mock SBP from the integral of $EM_{\rm true}$. After several iterations, the best-fitting parameters of the $\beta$-models are $n_{\rm 01}=(1.5\pm0.1) \times 10^{-3}$ cm$^{-3}$, $r_{\rm c1}=209.8\pm3.6$ kpc, $\beta_1=0.62\pm0.01$ for the SE subcluster centred on RA$=$11:44:50.1, DEC$=$+19:42:14.7; and $n_{\rm 02}=(7.1\pm0.2) \times 10^{-4}$ cm$^{-3}$, $r_{\rm c2}=218.3\pm8.4$ kpc, $\beta_2=0.57\pm0.02$ for the NW subcluster centred on RA$=$11:44:06.3, DEC$=$+19:54:58.9.
The total density is $n_{\rm e, ICM}=2.9\times10^{-4}{\rm\ cm}^{-3}$ for cluster gas around the OC (800 kpc from the SE subcluster centre and 450 kpc from the NW subcluster centre).
The average temperature is $kT_{\rm ICM}=2.9\pm0.2$ keV for the ICM in an annulus of 60-100 kpc around the OC.
Then the ICM thermal pressure is $P_{\rm ICM} = kn_{\rm tot}T_{\rm ICM}=kn_{\rm e,ICM}\cdot n_{\rm tot}/n_{\rm e}\cdot T_{\rm ICM}=2.6\times10^{-12}$ dyn cm$^{-2}$,
while the ram pressure from ICM is $P_{\rm ram}=\rho_{\rm ICM} v_{\rm OC}^2=5.6\times10^{-12}$ $(v_{\rm OC}/1000 {\rm\ km\ s}^{-1})^2$  dyn cm$^{-2}$.
This can be compared with the thermal pressure inside the OC, $P_{\rm OC}=1.5\times10^{-11}$ dyn cm$^{-2}$, from the average density.
Thus, the X-ray OC would be over-pressurized on sides not experiencing ram pressure, assuming a single $T$ for the OC. 
Including a density gradient in the OC can alleviate the pressure imbalance at the edge but the OC is still over-pressurized. Without an associated dark matter halo, the OC has to expand.

We can also examine the pressure balance assuming two phases of gas, with the 2{\tt APEC} fitting result in Table~\ref{t:fit} (the one with the same abundance for both phases).
The pressure ratio for two phases is $\frac{P_1}{P_2}=\frac{n_1T_1}{n_2T_2}=1$.
The normalization ratio of two {\tt APEC} model is $\frac{N_1}{N_2}=\frac{n_1^2f_1}{n_2^2f_2}$, 
where $f_1$ and $f_2$ are volume occupation factor for cool and hot phase gas and $f_1+f_2=f$.
Combining previous equations, we get
\begin{equation}
f_1=f\frac{N_1}{N_2}\Bigg/\left[\left(\frac{T_2}{T_1}\right)^2+\frac{N_1}{N_2}\right].
\end{equation}
The resultant $f_1=0.03f$ and $f_2=0.97f$, thus the hotter phase gas occupies 97 per cent of the volume of the soft X-ray emitting gas.
The electron density of cool and hot phase is  $n_{\rm e,cool}=8.5\times10^{-3}f^{-1/2}{\rm\ cm}^{-3}$ and $n_{\rm e,hot}=2.3\times10^{-3}f^{-1/2}{\rm\ cm}^{-3}$, respectively.
The total gas mass is $M_{\rm OC,2T}=1.2\times10^{10}f^{1/2}\ {\rm M}_\odot$. 
The ISM thermal pressure is $P_{\rm OC,2T}=2.6\times10^{-11}$ dyn cm$^{-2}$.
Thus, the pressure imbalance is even worse with the two-$T$ model.
There is a similar issue of pressure imbalance for stripped X-ray tails still attached to a galaxy (e.g. \citealt{2010ApJ...708..946S,2013ApJ...777..122Z}).
It is unclear whether pressure balance exists at the OC/ICM interface but possible solutions for the pressure imbalance include modeling uncertainty of the X-ray spectra (especially related to abundance), extra pressure support in the ICM from magnetic field and turbulence, and the contribution of charge exchange to the X-ray emission in stripped gas (e.g. \citealt{2013ApJ...777..122Z}).

\subsection{ICM microphysics of the OC}
The OC has been detached from the parent galaxy and the dark matter halo, it also presents an ideal example to study the ICM microphysics. 
How does it survive disruption by the Rayleigh-Taylor (RT) and Kelvin-Helmholtz (KH) instabilities, as well as the thermal conduction?

The KH instability occurs when there is a velocity difference across the interface between two fluids.
The typical mass loss rate due to KH instability is \citep{1982MNRAS.198.1007N}
\begin{equation}
\begin{split}
& \dot{M}_{\rm KH} \approx \pi R_{\rm OC}^2 \rho_{\rm ICM}v_{\rm OC} \\
& =23.9\left(\frac{n_{\rm e, ICM}}{2.9\times10^{-4}{\rm\ cm}^{-3}}\right)\left(\frac{R_{\rm OC}}{30{\rm\ kpc}}\right)^2\left(\frac{v_{\rm OC}}{1000 {\rm\ km\ s}^{-1}}\right){\rm M}_\odot {\rm\ yr}^{-1}.
\end{split}
\end{equation}
The mass loss timescale for the OC is $t_{\rm KH} = M_{\rm OC} / \dot M_{\rm KH} = 4.2 \times 10^8$ yr.
The KH instability can be suppressed by magnetic field if $\frac{B^2(\rho_1+\rho_2)}{2\pi\rho_1\rho_2 (v_1-v_2)^2} \geq 1$ \citep{1961hhs..book.....C}, where $B$ is the average tangential magnetic field of two fluids beside the interface.
In this case, $v_1-v_2=v_{\rm OC}$, $\rho_1=\rho_{\rm OC}$, $\rho_2=\rho_{\rm ICM}$, and $\rho_{\rm OC} \gg \rho_{\rm ICM}$, we assume $B=6~\mu G$, thus the condition for suppression of KH instability is 
\begin{equation}
\begin{split}
& \frac{B^2}{2\pi \rho_{\rm ICM} v_{\rm OC}^2} \\
& =1.02 \left(\frac{B}{6~\mu G}\right)^2 \left(\frac{n_{\rm e, ICM}}{2.9\times10^{-4}{\rm\ cm}^{-3}}\right)^{-1} \left(\frac{v_{\rm OC}}{1000 {\rm\ km\ s}^{-1}}\right)^{-2} \geq 1.
\end{split}
\end{equation}
The typical magnetic field is a few $\mu G$ (e.g. \citealt{2002ARA&A..40..319C,2010A&A...513A..30B}) in the ICM. However, magnetic fields can be amplified by cluster merge shocks (e.g. \citealt{2018SSRv..214..122D}), and OC is likely in the post-shock region. Moreover, the magnetic field near a moving cloud can be significantly strengthened by the formation of a parallel magnetic field layer via magnetic draping (e.g. \citealt{2008ApJ...677..993D,2021NatAs...5..159M}).
Thus, a magnetic field with a strength of $\sim 6~\mu G$ around the OC is possible and could suppress the KH instability.

The RT instability occurs in an interface between two fluids of different densities, when the lighter fluid is pushing the heavier one typically due to a gravitational field or an acceleration.
An equivalent situation applied here is the dense OC cloud moving through the rarefied ICM.
The drag force on the OC cloud is $F_{\rm d} = C_{\rm d} \rho_{\rm ICM}v_{\rm OC}^2 A/2$, where $C_{\rm d} \approx 0.5$ is the drag coefficient assuming a sphere shape for OC, $A=\pi R_{\rm OC}^2$ is the cloud cross-sectional area, and $\rho_{\rm ICM}v_{\rm OC}^2$ is the ram pressure $P_{\rm ram}$ from ICM.
The relevant acceleration is $a=F_{\rm d}/M_{\rm OC}=\frac{3C_{\rm d}P_{\rm ram}}{8\rho_{\rm OC}R_{\rm OC}} =1.9\times 10^{-9}$ cm s$^{-2}$, where $\rho_{\rm OC}$ is the OC density from the hot gas as the OC may not have an associated dark matter halo.
The RT instability would tear the cloud apart in a few characteristic e-folding times  
\begin{equation}
\begin{split}
& t_{\rm RT}=\left(\frac{\lambda}{2\pi a}\right)^{1/2} \\
& =8.8\times10^7 \left(\frac{\lambda}{30 {\rm\ kpc}}\right)^{1/2}\left(\frac{a}{1.9\times 10^{-9}{\rm\ cm\ s}^{-2}}\right)^{-1/2} {\rm\ yr},
\end{split}
\end{equation}
where $\lambda$ is the scale-length of the RT perturbation.
The RT instability can be stabilized by mechanisms such as self-gravity and magnetic fields \citep{1961hhs..book.....C}.
The self-gravitational acceleration of OC $g_{\rm OC}=\frac{GM_{\rm OC}}{R_{\rm OC}^2}=1.6\times 10^{-10}$ cm s$^{-2}$, which is much smaller than $a$;
thus the gas self-gravity is insufficient to suppress the RT instability.
The tension of magnetic field can suppress the growth of perturbations of scale-length $\lambda < \lambda_c$ with
\begin{equation}
\begin{split}
& \lambda_c=\frac{B^2{\rm cos}^2\theta}{a(\rho_{\rm OC}-\rho_{\rm ICM})} \\
& =511 \left(\frac{B}{6~\mu G}\right)^2 \left(\frac{a}{1.9\times 10^{-9}{\rm\ cm\ s}^{-2}}\right)^{-1} \left(\frac{n_{\rm e, OC}}{3.1\times 10^{-3}{\rm\ cm^{-3}}}\right)^{-1}\rm kpc,
\end{split}
\end{equation}
where an average value of $\langle{\rm cos}^2\theta\rangle= 1/2 $ is used and $\rho_{\rm OC} \gg \rho_{\rm ICM}$ thus  $\rho_{\rm ICM}$ is ignored here.
The $\lambda_c$ is 17 times larger than the radius of the OC cloud, thus a magnetic field around 6~$\mu G$ can also suppress the RT instability effectively.

The thermal conduction can smear out the temperature gradient between the OC and nearby ICM, i.e. the cooler OC evaporates in the hotter ICM. We can compare the size of the OC with a  critical length called `Field length' (e.g. \citealt{1990ApJ...358..392M}):
\begin{equation}
\begin{split}
& \lambda_{\rm F}=\left(\frac{\kappa T}{n^2\Lambda}\right)^{1/2}
=2.2\left(\frac{T_{\rm ICM}}{2.9 {\rm\ keV}}\right)^{7/4}\left(\frac{n_{\rm e,ICM}}{2.9 \times 10^{-4} {\rm\ cm}^{-3}}\right)^{-1} \\
& \times \left(\frac{\Lambda}{10^{-22.5}{\rm\ erg\ s}^{-1}{\rm cm}^3}\right)^{-1/2} {\rm\ Mpc},
\end{split}
\end{equation}
where $\kappa = 5.6 \times 10^{-7}T^{5/2}{\rm\ erg\ s}^{-1}{\rm\ K}^{-1}{\rm\ cm}^{-1}$ is the Spitzer conductivity \citep{1962pfig.book.....S}, $\Lambda$ is the X-ray cooling rate (e.g. \citealt{2009A&A...508..751S}).
The OC size is much smaller than the $\lambda_{\rm F}$, thus it will be evaporated by thermal conduction from the hot surrounding ICM on a conduction timescale (e.g. \citealt{1988xrec.book.....S}) of:
\begin{equation}
\begin{split}
& t_{\rm cond}=\frac{nkl^2}{\kappa} \\
& =2.9\times 10^7 \left(\frac{n_{\rm e,ICM}}{2.9 \times 10^{-4} {\rm\ cm}^{-3}}\right)\left(\frac{R_{\rm OC}}{30 {\rm\ kpc}}\right)^2\left(\frac{T_{\rm ICM}}{2.9{\rm\ keV}}\right)^{-5/2} {\rm yr}.
\end{split}
\end{equation}
The OC can only travel for a distance of $d=v_{\rm OC}\times t_{\rm cond}\sim 30 (v_{\rm OC}/1000 {\rm\ km\ s}^{-1})$ kpc, which is too short. 
The thermal conduction has to be suppressed significantly.
Previous studies suggest that magnetic field can help to suppress the thermal conduction by two orders of magnitude
relative to the classical Spitzer value, which is beneficial to the survival of OC in the ICM (e.g.  \citealt{2002ARA&A..40..319C,2007PhR...443....1M}).

\subsection{Excitation mechanism of the warm, ionized gas in OC}
The line diagnostics show LI(N)ER-like emission for the OC.
Several excitation mechanisms may produce LI(N)ER-like emission, such as photoionization by AGN and hot evolved stars,
radiative shocks, photoionization and thermal conduction from the hot (T $\gg$ 10$^{4}$ K) ICM
(\citealt{2008ARA&A..46..475H, 2012ApJ...747...61Y, 2019ARA&A..57..511K} and references therein). In the outer stripped tails of ESO 137-001,
emission-line flux ratios similar to those in A1367 OC have been found and explained as the results of photoionization (stripped ionized gas
or in situ \hii\ regions) plus radiative shocks \citep{2016MNRAS.455.2028F}. 
For A1367 OC, the slow radiative shock models \citep{2011ApJ...734...87R} are not able to reproduce the observed \OI/H$\alpha$
flux ratios. In addition, the median velocity dispersion of the ionized gas is only $\sim$ 80 \km, which is too low for shocks.

The OC has very weak SF at most. The {\em GALEX} data in this field, with 3953 s of exposure at the FUV and 4340 s of exposure at the NUV, are much deeper than the \XMM\ OM UVM2 data. The lack of any {\em GALEX} source in the {\em MUSE} field of the OC gives an upper limit on the SFR at 6$\times10^{-4}$ ${\rm M}_{\odot}{\rm\ yr}^{-1}$, with the calibration from \cite{2012ARA&A..50..531K}.
We also attempted to select \hii\ region candidates from the {\em MUSE} H$\alpha$ surface brightness map with SExtractor. By requesting point-like sources (CLASS\_STAR $>$ 0.9) with a low ellipticity ($e <$ 0.2),
only one candidate at RA=11:44:21.4 and DEC=+20:10:14.6 is identified. 
This candidate is also shown as the most compact clump in the \subaru\ net H$\alpha$ image. However, as shown in Fig.~\ref{fig:muse_maps}, it is off the main cloud and rather isolated.
As this source is too faint, we cannot unambiguously confirm it as an \hii\ region from its {\em MUSE} spectrum.
The H$\alpha$ luminosity of this source is 2.8$\times10^{37}$ erg s$^{-1}$ (without intrinsic extinction), which would
correspond to a SFR of 1.5$\times10^{-4}$ ${\rm M}_{\odot}{\rm\ yr}^{-1}$, with the calibration from \cite{2012ARA&A..50..531K}.
The lack of even weak SF in the OC excludes young stars as the main ionization source. However, there is SF ongoing in the whole cloud complex beyond the {\em MUSE} field, as shown in \cite{2017ApJ...839...65Y} and the two likely \hii\ regions observed with {\em APO/DIS}.

The models of photoionization from the hot ICM
\citep{1990ApJ...360L..15V,1991ApJ...381..361D} also have difficulty accounting for the observed emission-line flux ratios, especially the \OIII/H$\beta$ flux ratios.
\cite{2021arXiv210303147C} particularly used $kT \sim$ 1 keV plasma as the ionizing source in an RPS galaxy from GASP, but with the similar issues as before, also because the ionization parameter needs to be sufficiently large enough to account for the bulk of the observed optical line emission (or the line ratios are not the only constraints).
\cite{2009MNRAS.392.1475F} proposed collisional ionization (by cosmic rays and dissipation of magnetohydrodynamic wave energy)
to explain the filament emission in galaxy clusters. While the predicted \NII/H$\alpha$, \SII/H$\alpha$, and \OI/H$\alpha$ are close to our observational results,
the \OIII/H$\beta$ flux ratios are still under-predicted by at least two orders of magnitude. Further development of 
models and multi-wavelength diagnostics (i.e. infrared and ultraviolet lines) may help to explore the excitation mechanism of the LI(N)ER-like emission in A1367 OC.

While the metallicity of the warm, ionized gas would provide important information on its parent, the unclear ionization mechanism prevents a robust estimate of the metallicity. If we simply adopt the diagnostics for \hii\ regions, the derived metallicity is log(O/H) + 12 = 8.5 - 8.9 from the \NII $\lambda$6584 / \SII $\lambda\lambda$ 6716, 6731 and the \NII $\lambda$6584 / H$\alpha$ ratios, with the diagnostics derived by \cite{2002ApJS..142...35K}. This about solar metallicity would imply a massive parent for the OC.
For the only \hii\ region candidate, the same diagnostics result in a metallicity of 8.6 - 8.8, consistent with the above estimate.

\subsection{A signpost of ICM clumping and turbulence}
After being stripped far away from the parent galaxy, now the OC is a clump in the ICM of A1367.
The clumpiness of the ICM has been studied with X-ray observations from the surface brightness fluctuations (e.g. \citealt{2012MNRAS.421.1123C, 2017MNRAS.469.2423M}) or ICM radial profiles (e.g. \citealt{2011Sci...331.1576S,2013MNRAS.432..554W,2015MNRAS.447.2198E}), especially at the cluster outskirt, because the gas clumping factor increases with the cluster radius suggested by simulations (e.g. \citealt{2011ApJ...731L..10N,2013MNRAS.429..799V}).
For example, \cite{2013MNRAS.429..799V} found that the typical X-ray clump size is 
$< 69$ kpc and the typical bolometric luminosity is $L_{\rm bol}$ = 4 $\times 10^{39} - 1.5 \times 10^{42} {\rm\ ergs\ s}^{-1}$.
The overall properties of the OC are consistent with the predicted properties of large/luminous ICM clumps from simulations. As a signpost of the ICM clumping in the nearby cluster A1367, we have a rare opportunity to study an ICM clump in detail.

Apart from providing additional non-thermal pressure to balance gravity, the ICM turbulence can also reaccelerate relativistic electrons and amplify the magnetic field to produce radio halo emission (e.g. \citealt{2003ApJ...584..190F,2007MNRAS.378..245B,2016ApJ...817..127B,2018SSRv..214..122D}); distribute energy and metals from AGN and stellar feedback (e.g. \citealt{2006MNRAS.372.1840R,2012ApJ...746...94G,2014Natur.515...85Z}); dissolve the ISM of galaxies through turbulent viscous stripping (e.g. \citealt{2005A&A...433..875R}).
The turbulence is generated from cluster merger or accretion of matter, cool core sloshing, and jet outflows from AGN (e.g. \citealt{2012A&A...544A.103V}).
The measurements of turbulence in cluster central regions directly from line width (e.g. \citealt{2010MNRAS.402L..11S,2016Natur.535..117H}) or indirectly from density/pressure fluctuation (e.g. \citealt{2004A&A...426..387S,2012MNRAS.421.1123C}) suggest that the ratio of turbulent pressure to thermal pressure in the ICM is small, i.e. $\lesssim 10$ per cent.
As for the clumping factor, simulations (e.g. \citealt{2009ApJ...705.1129L}) also suggest that the turbulence increases with cluster radius. Although the indirect measurements (e.g. \citealt{2016MNRAS.463..655K,2019A&A...621A..40E}) support this trend, there is limited observational evidence from direct measurements.

The multi-phase nature of OC provides us an opportunity to study the ICM turbulence at the cluster outskirt. As warm gas mixing with hot gas, the kinematics of multi-phase gas are tightly linked, such a connection has been proposed in cool cores \citep{2018ApJ...854..167G}.
The velocity dispersion of the warm gas is rather low, $\sim 80$ \km\ (Fig.~\ref{fig:muse_maps}),
which is much smaller than the sound speed in the hot gas (650 - 870 \km\ for $kT$ = 1.6 - 3.0 keV).
The ratio of turbulent pressure to the ICM thermal pressure in the OC is only $\sim 1$ per cent, if the kinematics of the warm ionized gas trace the kinematics of the ICM in this case.
Moreover, such a level of turbulence may be able to induce the slow top-down turbulent condensation from the ICM as suggested by the condensation criterion C-ratio $C \equiv t_{\rm cool}/t_{\rm eddy}$, where $t_{\rm eddy}=2 \pi r^{2/3}L^{1/3}/\sigma_{\rm v,L}$ is the turbulence eddy turnover time \citep{2018ApJ...854..167G}. In the case of the OC, we use a radius $r=30$ kpc, or a full size of cloud $L=60$ kpc, and a mean three-dimensional velocity dispersion $\sigma_{\rm v,L}=\sqrt{3}\sigma_{\rm v,1d}=139$ \km. We then have $t_{\rm cool}=3.6-7.2$ Gyr and $t_{\rm eddy}=1.6 $ Gyr that returns a $C=2.3-4.5$, which indicates the condensation may be significant over the long term. Therefore, the OC may grow via slow ICM condensation and accretion, akin to an off-centre chaotic cold accretion rain \citep{2017MNRAS.466..677G}.
Such kind of growth for intracluster clouds is also suggested in the simulation (e.g. \citealt{2020MNRAS.499.4261S,2021MNRAS.501.1143K}).

\section{Conclusion}
We have discovered an OC detected in both H$\alpha$ and X-rays in the outskirt of the merging galaxy cluster A1367, with a projected distance of $\sim 800$ kpc to the cluster centre. Our main conclusions are as follows:

\begin{itemize}
\item [1)] The cloud most likely originates from the stripped ISM of an infalling galaxy. The parent galaxy is still unknown and maybe a massive one, because the OC has an X-ray bolometric luminosity of $\sim 3.1\times 10^{41}$ erg s$^{-1}$ and a hot gas mass of $\sim 10^{10}\ {\rm M}_\odot$. The metallicity of the H$\alpha$ OC also suggests a massive parent galaxy.

\item [2)] By being stripped and far away from the parent galaxy, now the OC is a signpost of the ICM clumping. It may be in an advanced evolutionary stage suggested by a higher average X-ray temperature of 1.6 keV than typical X-ray tails and generally high X-ray-to-H$\alpha$ ratios than stripped tails still attached to their parent galaxies.

\item [3)] The H$\alpha$ peak of the OC has an offset of $\sim$ 12 kpc from the X-ray peak of the OC, with several H$\alpha$ filaments enclosing the X-ray peak. The bright H$\alpha$ clumps may be associated with the only surviving cold clouds mixing with the ICM. 

\item [4)] The H$\alpha$ OC shows a velocity gradient along the east-west direction as an indication of cloud's motion, but with a low level of velocity dispersion ($\sim$ 80 \km) likely indicating a low level of the ICM turbulence. 

\item [5)] The line diagnostics from \muse\ suggest little SF in the main H$\alpha$ OC and a LI(N)ER-like spectrum.
The non-detection of {\em GALEX} source in the {\em MUSE} field of the OC gives an upper limit on the SFR at $\sim 10^{-3}$\ ${\rm M}_{\odot}{\rm\ yr}^{-1}$ for the main body of the H$\alpha$ OC, but some SF is present to the SE of OC.

\item [6)] It is found that a magnetic field around $6~\mu$G can suppress hydrodynamic instabilities (RT and KH instabilities) and thermal conduction to help the survival of the cloud in the harsh ICM environment.
\end{itemize}

This discovery of an isolated X-ray clump accompanied by H$\alpha$ emission suggests that some ICM clumps are multi-phase.
Future multi-wavelength observations can explore the multi-phase nature of ICM clumps better, and link them to the related multi-phase processes of cool cores (e.g. \citealt{2020NatAs...4...10G}). Moreover, this discovery suggests that we can potentially probe ICM clumping with future sensitive and wide-field H$\alpha$ surveys (e.g. \citealt{2018A&A...614A..56B}). 
The kinematics of the ICM may also be explored with warm gas in the future.

While the OC is only the first ICM clump detected in both X-rays and H$\alpha$, and ICM clumps as luminous as the OC may be rare, the number of similar examples should grow with more sensitive X-ray (e.g. {\em eROSITA}) and H$\alpha$ data to survey nearby galaxy clusters. ICM clumps as luminous as the A1367 OC can be detected with 30 ks clean \XMM\ time up to $z=0.056$. Similarly, with 80 ks clean \cha\ ACIS-I time, we can detect the same clump out to $z=0.050$ ($z=0.039$) with the \cha\ cycle 5 (cycle 23) response. The detection does depend on the local ICM background. If the local ICM background is increased by a factor of 3, 30 ks clean \XMM\ observation can only detect the OC up to $z=0.048$.
H$\alpha$ clouds like the A1367 OC can also be detected to the above $z$ with the similar H$\alpha$ narrow-band imaging data as those in \cite{2017ApJ...839...65Y}. The multi-wavelength surveys and studies will be necessary for us to better understand the multi-phase isolated clouds in clusters.

\section*{Acknowledgements}
Support for this work was provided by the NASA grants 80NSSC19K0953 and 80NSSC19K1257 and the NSF grant 1714764.
M.F. has received funding from the European Research Council (ERC) under the European Union's Horizon 2020 research and innovation programme (grant agreement No 757535).
M.G. acknowledges partial support by NASA Chandra GO8-19104X/GO9-20114X and HST GO-15890.020-A.
J.K. is supported by NSF through grant AST-1812847.
This research is based on observations obtained with \xmm, an ESA science mission with instruments and contributions directly funded by ESA Member States and NASA.
This research is also based on observations collected at the European Southern Observatory under ESO programme 0104.A-0268(A). This research is also based on observations with the \subaru\ telescope and observations obtained with the {\em APO} 3.5-meter telescope, which is owned and operated by the Astrophysical Research Consortium.
We appreciate the support from the F. H. Levinson Fund of the Silicon Valley Community Foundation.
Data analysis was in part carried out on the Multi-wavelength Data Analysis System operated by the Astronomy Data Center (ADC), National Astronomical Observatory of Japan.

\section*{DATA AVAILABILITY}
The \xmm\ raw data used in this paper are available to download at the HEASARC Data Archive website\footnote{https://heasarc.gsfc.nasa.gov/docs/archive.html}.
The \muse\ raw data are available to download at the ESO Science Archive Facility\footnote{http://archive.eso.org/cms.html}.
The reduced data underlying this paper will be shared on reasonable requests to the corresponding authors.

\bibliographystyle{mnras}
\bibliography{ms.bib}

\begin{thebibliography}{}
\makeatletter
\relax
\def\mn@urlcharsother{\let\do\@makeother \do\$\do\&\do\#\do\^\do\_\do\%\do\~}
\def\mn@doi{\begingroup\mn@urlcharsother \@ifnextchar [ {\mn@doi@}
  {\mn@doi@[]}}
\def\mn@doi@[#1]#2{\def\@tempa{#1}\ifx\@tempa\@empty \href
  {http://dx.doi.org/#2} {doi:#2}\else \href {http://dx.doi.org/#2} {#1}\fi
  \endgroup}
\def\mn@eprint#1#2{\mn@eprint@#1:#2::\@nil}
\def\mn@eprint@arXiv#1{\href {http://arxiv.org/abs/#1} {{\tt arXiv:#1}}}
\def\mn@eprint@dblp#1{\href {http://dblp.uni-trier.de/rec/bibtex/#1.xml}
  {dblp:#1}}
\def\mn@eprint@#1:#2:#3:#4\@nil{\def\@tempa {#1}\def\@tempb {#2}\def\@tempc
  {#3}\ifx \@tempc \@empty \let \@tempc \@tempb \let \@tempb \@tempa \fi \ifx
  \@tempb \@empty \def\@tempb {arXiv}\fi \@ifundefined
  {mn@eprint@\@tempb}{\@tempb:\@tempc}{\expandafter \expandafter \csname
  mn@eprint@\@tempb\endcsname \expandafter{\@tempc}}}

\bibitem[\protect\citeauthoryear{{Asplund}, {Grevesse}, {Sauval}  \&
  {Scott}}{{Asplund} et~al.}{2009}]{2009ARA&A..47..481A}
{Asplund} M.,  {Grevesse} N.,  {Sauval} A.~J.,   {Scott} P.,  2009, \mn@doi
  [\araa] {10.1146/annurev.astro.46.060407.145222}, \href
  {https://ui.adsabs.harvard.edu/abs/2009ARA&A..47..481A} {47, 481}

\bibitem[\protect\citeauthoryear{{Bacon} et~al.,}{{Bacon}
  et~al.}{2010}]{2010SPIE.7735E..08B}
{Bacon} R.,  et~al., 2010, in {McLean} I.~S.,  {Ramsay} S.~K.,   {Takami} H.,
  eds,  Society of Photo-Optical Instrumentation Engineers (SPIE) Conference
  Series Vol. 7735, Ground-based and Airborne Instrumentation for Astronomy
  III. p. 773508, \mn@doi{10.1117/12.856027}

\bibitem[\protect\citeauthoryear{{Bacon}, {Piqueras}, {Conseil}, {Richard}  \&
  {Shepherd}}{{Bacon} et~al.}{2016}]{2016ascl.soft11003B}
{Bacon} R.,  {Piqueras} L.,  {Conseil} S.,  {Richard} J.,   {Shepherd} M.,
  2016, {MPDAF: MUSE Python Data Analysis Framework}

\bibitem[\protect\citeauthoryear{{Beccari} et~al.,}{{Beccari}
  et~al.}{2017}]{2017MNRAS.465.2189B}
{Beccari} G.,  et~al., 2017, \mn@doi [\mnras] {10.1093/mnras/stw2874}, \href
  {https://ui.adsabs.harvard.edu/abs/2017MNRAS.465.2189B} {465, 2189}

\bibitem[\protect\citeauthoryear{{Bellazzini} et~al.,}{{Bellazzini}
  et~al.}{2018}]{2018MNRAS.476.4565B}
{Bellazzini} M.,  et~al., 2018, \mn@doi [\mnras] {10.1093/mnras/sty467}, \href
  {https://ui.adsabs.harvard.edu/abs/2018MNRAS.476.4565B} {476, 4565}

\bibitem[\protect\citeauthoryear{{Beresnyak} \& {Miniati}}{{Beresnyak} \&
  {Miniati}}{2016}]{2016ApJ...817..127B}
{Beresnyak} A.,  {Miniati} F.,  2016, \mn@doi [\apj]
  {10.3847/0004-637X/817/2/127}, \href
  {https://ui.adsabs.harvard.edu/abs/2016ApJ...817..127B} {817, 127}

\bibitem[\protect\citeauthoryear{{Bertin}, {Mellier}, {Radovich}, {Missonnier},
  {Didelon}  \& {Morin}}{{Bertin} et~al.}{2002}]{2002ASPC..281..228B}
{Bertin} E.,  {Mellier} Y.,  {Radovich} M.,  {Missonnier} G.,  {Didelon} P.,
  {Morin} B.,  2002, in {Bohlender} D.~A.,  {Durand} D.,   {Handley} T.~H.,
  eds,  Astronomical Society of the Pacific Conference Series Vol. 281,
  Astronomical Data Analysis Software and Systems XI. p.~228

\bibitem[\protect\citeauthoryear{{Bonafede}, {Feretti}, {Murgia}, {Govoni},
  {Giovannini}, {Dallacasa}, {Dolag}  \& {Taylor}}{{Bonafede}
  et~al.}{2010}]{2010A&A...513A..30B}
{Bonafede} A.,  {Feretti} L.,  {Murgia} M.,  {Govoni} F.,  {Giovannini} G.,
  {Dallacasa} D.,  {Dolag} K.,   {Taylor} G.~B.,  2010, \mn@doi [\aap]
  {10.1051/0004-6361/200913696}, \href
  {https://ui.adsabs.harvard.edu/abs/2010A&A...513A..30B} {513, A30}

\bibitem[\protect\citeauthoryear{{Bosch} et~al.,}{{Bosch}
  et~al.}{2018}]{2018PASJ...70S...5B}
{Bosch} J.,  et~al., 2018, \mn@doi [\pasj] {10.1093/pasj/psx080}, \href
  {https://ui.adsabs.harvard.edu/abs/2018PASJ...70S...5B} {70, S5}

\bibitem[\protect\citeauthoryear{{Boselli} \& {Gavazzi}}{{Boselli} \&
  {Gavazzi}}{2006}]{2006PASP..118..517B}
{Boselli} A.,  {Gavazzi} G.,  2006, \mn@doi [\pasp] {10.1086/500691}, \href
  {https://ui.adsabs.harvard.edu/abs/2006PASP..118..517B} {118, 517}

\bibitem[\protect\citeauthoryear{{Boselli} et~al.,}{{Boselli}
  et~al.}{2016}]{2016A&A...587A..68B}
{Boselli} A.,  et~al., 2016, \mn@doi [\aap] {10.1051/0004-6361/201527795},
  \href {https://ui.adsabs.harvard.edu/abs/2016A&A...587A..68B} {587, A68}

\bibitem[\protect\citeauthoryear{{Boselli} et~al.,}{{Boselli}
  et~al.}{2018}]{2018A&A...614A..56B}
{Boselli} A.,  et~al., 2018, \mn@doi [\aap] {10.1051/0004-6361/201732407},
  \href {https://ui.adsabs.harvard.edu/abs/2018A&A...614A..56B} {614, A56}

\bibitem[\protect\citeauthoryear{{Boselli} et~al.,}{{Boselli}
  et~al.}{2021}]{2021A&A...646A.139B}
{Boselli} A.,  et~al., 2021, \mn@doi [\aap] {10.1051/0004-6361/202039046},
  \href {https://ui.adsabs.harvard.edu/abs/2021A&A...646A.139B} {646, A139}

\bibitem[\protect\citeauthoryear{{Brunetti} \& {Lazarian}}{{Brunetti} \&
  {Lazarian}}{2007}]{2007MNRAS.378..245B}
{Brunetti} G.,  {Lazarian} A.,  2007, \mn@doi [\mnras]
  {10.1111/j.1365-2966.2007.11771.x}, \href
  {https://ui.adsabs.harvard.edu/abs/2007MNRAS.378..245B} {378, 245}

\bibitem[\protect\citeauthoryear{{Burkhart} \& {Loeb}}{{Burkhart} \&
  {Loeb}}{2016}]{2016ApJ...824L...7B}
{Burkhart} B.,  {Loeb} A.,  2016, \mn@doi [\apjl] {10.3847/2041-8205/824/1/L7},
  \href {https://ui.adsabs.harvard.edu/abs/2016ApJ...824L...7B} {824, L7}

\bibitem[\protect\citeauthoryear{{Calura}, {Bellazzini}  \&
  {D'Ercole}}{{Calura} et~al.}{2020}]{2020MNRAS.499.5873C}
{Calura} F.,  {Bellazzini} M.,   {D'Ercole} A.,  2020, \mn@doi [\mnras]
  {10.1093/mnras/staa3133}, \href
  {https://ui.adsabs.harvard.edu/abs/2020MNRAS.499.5873C} {499, 5873}

\bibitem[\protect\citeauthoryear{{Campitiello} et~al.,}{{Campitiello}
  et~al.}{2021}]{2021arXiv210303147C}
{Campitiello} M.~G.,  et~al., 2021, arXiv e-prints, \href
  {https://ui.adsabs.harvard.edu/abs/2021arXiv210303147C} {p. arXiv:2103.03147}

\bibitem[\protect\citeauthoryear{{Cannon} et~al.,}{{Cannon}
  et~al.}{2015}]{2015AJ....149...72C}
{Cannon} J.~M.,  et~al., 2015, \mn@doi [\aj] {10.1088/0004-6256/149/2/72},
  \href {https://ui.adsabs.harvard.edu/abs/2015AJ....149...72C} {149, 72}

\bibitem[\protect\citeauthoryear{{Carilli} \& {Taylor}}{{Carilli} \&
  {Taylor}}{2002}]{2002ARA&A..40..319C}
{Carilli} C.~L.,  {Taylor} G.~B.,  2002, \mn@doi [\araa]
  {10.1146/annurev.astro.40.060401.093852}, \href
  {https://ui.adsabs.harvard.edu/abs/2002ARA&A..40..319C} {40, 319}

\bibitem[\protect\citeauthoryear{{Cavaliere} \& {Fusco-Femiano}}{{Cavaliere} \&
  {Fusco-Femiano}}{1976}]{1976A&A....49..137C}
{Cavaliere} A.,  {Fusco-Femiano} R.,  1976, \aap, \href
  {https://ui.adsabs.harvard.edu/abs/1976A&A....49..137C} {500, 95}

\bibitem[\protect\citeauthoryear{{Chandrasekhar}}{{Chandrasekhar}}{1961}]{1961hhs..book.....C}
{Chandrasekhar} S.,  1961, {Hydrodynamic and hydromagnetic stability}

\bibitem[\protect\citeauthoryear{{Chen} et~al.,}{{Chen}
  et~al.}{2020}]{2020MNRAS.496.4654C}
{Chen} H.,  et~al., 2020, \mn@doi [\mnras] {10.1093/mnras/staa1868}, \href
  {https://ui.adsabs.harvard.edu/abs/2020MNRAS.496.4654C} {496, 4654}

\bibitem[\protect\citeauthoryear{{Chung}, {van Gorkom}, {Kenney}  \&
  {Vollmer}}{{Chung} et~al.}{2007}]{2007ApJ...659L.115C}
{Chung} A.,  {van Gorkom} J.~H.,  {Kenney} J. D.~P.,   {Vollmer} B.,  2007,
  \mn@doi [\apjl] {10.1086/518034}, \href
  {https://ui.adsabs.harvard.edu/abs/2007ApJ...659L.115C} {659, L115}

\bibitem[\protect\citeauthoryear{{Churazov} et~al.,}{{Churazov}
  et~al.}{2012}]{2012MNRAS.421.1123C}
{Churazov} E.,  et~al., 2012, \mn@doi [\mnras]
  {10.1111/j.1365-2966.2011.20372.x}, \href
  {https://ui.adsabs.harvard.edu/abs/2012MNRAS.421.1123C} {421, 1123}

\bibitem[\protect\citeauthoryear{{Cid Fernandes}, {Stasi{\'n}ska},
  {Schlickmann}, {Mateus}, {Vale Asari}, {Schoenell}  \& {Sodr{\'e}}}{{Cid
  Fernandes} et~al.}{2010}]{2010MNRAS.403.1036C}
{Cid Fernandes} R.,  {Stasi{\'n}ska} G.,  {Schlickmann} M.~S.,  {Mateus} A.,
  {Vale Asari} N.,  {Schoenell} W.,   {Sodr{\'e}} L.,  2010, \mn@doi [\mnras]
  {10.1111/j.1365-2966.2009.16185.x}, \href
  {https://ui.adsabs.harvard.edu/abs/2010MNRAS.403.1036C} {403, 1036}

\bibitem[\protect\citeauthoryear{{Consolandi}, {Gavazzi}, {Fossati},
  {Fumagalli}, {Boselli}, {Yagi}  \& {Yoshida}}{{Consolandi}
  et~al.}{2017}]{2017A&A...606A..83C}
{Consolandi} G.,  {Gavazzi} G.,  {Fossati} M.,  {Fumagalli} M.,  {Boselli} A.,
  {Yagi} M.,   {Yoshida} M.,  2017, \mn@doi [\aap]
  {10.1051/0004-6361/201731218}, \href
  {https://ui.adsabs.harvard.edu/abs/2017A&A...606A..83C} {606, A83}

\bibitem[\protect\citeauthoryear{{Cortese}, {Gavazzi}, {Boselli},
  {Iglesias-Paramo}  \& {Carrasco}}{{Cortese}
  et~al.}{2004}]{2004A&A...425..429C}
{Cortese} L.,  {Gavazzi} G.,  {Boselli} A.,  {Iglesias-Paramo} J.,   {Carrasco}
  L.,  2004, \mn@doi [\aap] {10.1051/0004-6361:20040381}, \href
  {https://ui.adsabs.harvard.edu/abs/2004A&A...425..429C} {425, 429}

\bibitem[\protect\citeauthoryear{{Davies} et~al.,}{{Davies}
  et~al.}{2004}]{2004MNRAS.349..922D}
{Davies} J.,  et~al., 2004, \mn@doi [\mnras]
  {10.1111/j.1365-2966.2004.07568.x}, \href
  {https://ui.adsabs.harvard.edu/abs/2004MNRAS.349..922D} {349, 922}

\bibitem[\protect\citeauthoryear{{Dolag}, {Borgani}, {Murante}  \&
  {Springel}}{{Dolag} et~al.}{2009}]{2009MNRAS.399..497D}
{Dolag} K.,  {Borgani} S.,  {Murante} G.,   {Springel} V.,  2009, \mn@doi
  [\mnras] {10.1111/j.1365-2966.2009.15034.x}, \href
  {https://ui.adsabs.harvard.edu/abs/2009MNRAS.399..497D} {399, 497}

\bibitem[\protect\citeauthoryear{{Donahue} \& {Voit}}{{Donahue} \&
  {Voit}}{1991}]{1991ApJ...381..361D}
{Donahue} M.,  {Voit} G.~M.,  1991, \mn@doi [\apj] {10.1086/170659}, \href
  {https://ui.adsabs.harvard.edu/abs/1991ApJ...381..361D} {381, 361}

\bibitem[\protect\citeauthoryear{{Donnert}, {Vazza}, {Br{\"u}ggen}  \&
  {ZuHone}}{{Donnert} et~al.}{2018}]{2018SSRv..214..122D}
{Donnert} J.,  {Vazza} F.,  {Br{\"u}ggen} M.,   {ZuHone} J.,  2018, \mn@doi
  [\ssr] {10.1007/s11214-018-0556-8}, \href
  {https://ui.adsabs.harvard.edu/abs/2018SSRv..214..122D} {214, 122}

\bibitem[\protect\citeauthoryear{{Duc} \& {Bournaud}}{{Duc} \&
  {Bournaud}}{2008}]{2008ApJ...673..787D}
{Duc} P.-A.,  {Bournaud} F.,  2008, \mn@doi [\apj] {10.1086/524868}, \href
  {https://ui.adsabs.harvard.edu/abs/2008ApJ...673..787D} {673, 787}

\bibitem[\protect\citeauthoryear{{Dursi} \& {Pfrommer}}{{Dursi} \&
  {Pfrommer}}{2008}]{2008ApJ...677..993D}
{Dursi} L.~J.,  {Pfrommer} C.,  2008, \mn@doi [\apj] {10.1086/529371}, \href
  {https://ui.adsabs.harvard.edu/abs/2008ApJ...677..993D} {677, 993}

\bibitem[\protect\citeauthoryear{{ESO CPL Development Team}}{{ESO CPL
  Development Team}}{2015}]{2015ascl.soft04003E}
{ESO CPL Development Team} 2015, {EsoRex: ESO Recipe Execution Tool}

\bibitem[\protect\citeauthoryear{{Eckert}, {Roncarelli}, {Ettori}, {Molendi},
  {Vazza}, {Gastaldello}  \& {Rossetti}}{{Eckert}
  et~al.}{2015}]{2015MNRAS.447.2198E}
{Eckert} D.,  {Roncarelli} M.,  {Ettori} S.,  {Molendi} S.,  {Vazza} F.,
  {Gastaldello} F.,   {Rossetti} M.,  2015, \mn@doi [\mnras]
  {10.1093/mnras/stu2590}, \href
  {https://ui.adsabs.harvard.edu/abs/2015MNRAS.447.2198E} {447, 2198}

\bibitem[\protect\citeauthoryear{{Eckert} et~al.,}{{Eckert}
  et~al.}{2019}]{2019A&A...621A..40E}
{Eckert} D.,  et~al., 2019, \mn@doi [\aap] {10.1051/0004-6361/201833324}, \href
  {https://ui.adsabs.harvard.edu/abs/2019A&A...621A..40E} {621, A40}

\bibitem[\protect\citeauthoryear{{Ferland}, {Fabian}, {Hatch}, {Johnstone},
  {Porter}, {van Hoof}  \& {Williams}}{{Ferland}
  et~al.}{2009}]{2009MNRAS.392.1475F}
{Ferland} G.~J.,  {Fabian} A.~C.,  {Hatch} N.~A.,  {Johnstone} R.~M.,  {Porter}
  R.~L.,  {van Hoof} P.~A.~M.,   {Williams} R.~J.~R.,  2009, \mn@doi [\mnras]
  {10.1111/j.1365-2966.2008.14153.x}, \href
  {https://ui.adsabs.harvard.edu/abs/2009MNRAS.392.1475F} {392, 1475}

\bibitem[\protect\citeauthoryear{{Fitzpatrick}}{{Fitzpatrick}}{1999}]{1999PASP..111...63F}
{Fitzpatrick} E.~L.,  1999, \mn@doi [\pasp] {10.1086/316293}, \href
  {https://ui.adsabs.harvard.edu/abs/1999PASP..111...63F} {111, 63}

\bibitem[\protect\citeauthoryear{{Fossati}, {Fumagalli}, {Boselli}, {Gavazzi},
  {Sun}  \& {Wilman}}{{Fossati} et~al.}{2016}]{2016MNRAS.455.2028F}
{Fossati} M.,  {Fumagalli} M.,  {Boselli} A.,  {Gavazzi} G.,  {Sun} M.,
  {Wilman} D.~J.,  2016, \mn@doi [\mnras] {10.1093/mnras/stv2400}, \href
  {https://ui.adsabs.harvard.edu/abs/2016MNRAS.455.2028F} {455, 2028}

\bibitem[\protect\citeauthoryear{{Fossati}, {Fumagalli}, {Gavazzi},
  {Consolandi}, {Boselli}, {Yagi}, {Sun}  \& {Wilman}}{{Fossati}
  et~al.}{2019}]{2019MNRAS.484.2212F}
{Fossati} M.,  {Fumagalli} M.,  {Gavazzi} G.,  {Consolandi} G.,  {Boselli} A.,
  {Yagi} M.,  {Sun} M.,   {Wilman} D.~J.,  2019, \mn@doi [\mnras]
  {10.1093/mnras/stz136}, \href
  {https://ui.adsabs.harvard.edu/abs/2019MNRAS.484.2212F} {484, 2212}

\bibitem[\protect\citeauthoryear{{Fujita}, {Takizawa}  \& {Sarazin}}{{Fujita}
  et~al.}{2003}]{2003ApJ...584..190F}
{Fujita} Y.,  {Takizawa} M.,   {Sarazin} C.~L.,  2003, \mn@doi [\apj]
  {10.1086/345599}, \href
  {https://ui.adsabs.harvard.edu/abs/2003ApJ...584..190F} {584, 190}

\bibitem[\protect\citeauthoryear{{Gaspari}, {Ruszkowski}  \&
  {Sharma}}{{Gaspari} et~al.}{2012}]{2012ApJ...746...94G}
{Gaspari} M.,  {Ruszkowski} M.,   {Sharma} P.,  2012, \mn@doi [\apj]
  {10.1088/0004-637X/746/1/94}, \href
  {https://ui.adsabs.harvard.edu/abs/2012ApJ...746...94G} {746, 94}

\bibitem[\protect\citeauthoryear{{Gaspari}, {Temi}  \& {Brighenti}}{{Gaspari}
  et~al.}{2017}]{2017MNRAS.466..677G}
{Gaspari} M.,  {Temi} P.,   {Brighenti} F.,  2017, \mn@doi [\mnras]
  {10.1093/mnras/stw3108}, \href
  {https://ui.adsabs.harvard.edu/abs/2017MNRAS.466..677G} {466, 677}

\bibitem[\protect\citeauthoryear{{Gaspari} et~al.,}{{Gaspari}
  et~al.}{2018}]{2018ApJ...854..167G}
{Gaspari} M.,  et~al., 2018, \mn@doi [\apj] {10.3847/1538-4357/aaaa1b}, \href
  {https://ui.adsabs.harvard.edu/abs/2018ApJ...854..167G} {854, 167}

\bibitem[\protect\citeauthoryear{{Gaspari}, {Tombesi}  \& {Cappi}}{{Gaspari}
  et~al.}{2020}]{2020NatAs...4...10G}
{Gaspari} M.,  {Tombesi} F.,   {Cappi} M.,  2020, \mn@doi [Nature Astronomy]
  {10.1038/s41550-019-0970-1}, \href
  {https://ui.adsabs.harvard.edu/abs/2020NatAs...4...10G} {4, 10}

\bibitem[\protect\citeauthoryear{{Gavazzi}, {Boselli}, {Mayer},
  {Iglesias-Paramo}, {V{\'\i}lchez}  \& {Carrasco}}{{Gavazzi}
  et~al.}{2001}]{2001ApJ...563L..23G}
{Gavazzi} G.,  {Boselli} A.,  {Mayer} L.,  {Iglesias-Paramo} J.,
  {V{\'\i}lchez} J.~M.,   {Carrasco} L.,  2001, \mn@doi [\apjl]
  {10.1086/338389}, \href
  {https://ui.adsabs.harvard.edu/abs/2001ApJ...563L..23G} {563, L23}

\bibitem[\protect\citeauthoryear{{Gavazzi}, {Consolandi}, {Yagi}  \&
  {Yoshida}}{{Gavazzi} et~al.}{2017}]{2017A&A...606A.131G}
{Gavazzi} G.,  {Consolandi} G.,  {Yagi} M.,   {Yoshida} M.,  2017, \mn@doi
  [\aap] {10.1051/0004-6361/201731372}, \href
  {https://ui.adsabs.harvard.edu/abs/2017A&A...606A.131G} {606, A131}

\bibitem[\protect\citeauthoryear{{Ge}, {Wang}, {Tripp}, {Li}, {Gu}  \&
  {Ji}}{{Ge} et~al.}{2016}]{2016MNRAS.459..366G}
{Ge} C.,  {Wang} Q.~D.,  {Tripp} T.~M.,  {Li} Z.,  {Gu} Q.,   {Ji} L.,  2016,
  \mn@doi [\mnras] {10.1093/mnras/stw599}, \href
  {https://ui.adsabs.harvard.edu/abs/2016MNRAS.459..366G} {459, 366}

\bibitem[\protect\citeauthoryear{{Ge}, {Sun}, {Rozo}, {Sehgal}, {Vikhlinin},
  {Forman}, {Jones}  \& {Nagai}}{{Ge} et~al.}{2019a}]{2019MNRAS.484.1946G}
{Ge} C.,  {Sun} M.,  {Rozo} E.,  {Sehgal} N.,  {Vikhlinin} A.,  {Forman} W.,
  {Jones} C.,   {Nagai} D.,  2019a, \mn@doi [\mnras] {10.1093/mnras/stz088},
  \href {https://ui.adsabs.harvard.edu/abs/2019MNRAS.484.1946G} {484, 1946}

\bibitem[\protect\citeauthoryear{{Ge} et~al.,}{{Ge}
  et~al.}{2019b}]{2019MNRAS.486L..36G}
{Ge} C.,  et~al., 2019b, \mn@doi [\mnras] {10.1093/mnrasl/slz049}, \href
  {https://ui.adsabs.harvard.edu/abs/2019MNRAS.486L..36G} {486, L36}

\bibitem[\protect\citeauthoryear{{Ghizzardi}, {Rossetti}  \&
  {Molendi}}{{Ghizzardi} et~al.}{2010}]{2010A&A...516A..32G}
{Ghizzardi} S.,  {Rossetti} M.,   {Molendi} S.,  2010, \mn@doi [\aap]
  {10.1051/0004-6361/200912496}, \href
  {https://ui.adsabs.harvard.edu/abs/2010A&A...516A..32G} {516, A32}

\bibitem[\protect\citeauthoryear{{Giles} et~al.,}{{Giles}
  et~al.}{2016}]{2016A&A...592A...3G}
{Giles} P.~A.,  et~al., 2016, \mn@doi [\aap] {10.1051/0004-6361/201526886},
  \href {https://ui.adsabs.harvard.edu/abs/2016A&A...592A...3G} {592, A3}

\bibitem[\protect\citeauthoryear{{Gunn} \& {Gott}}{{Gunn} \&
  {Gott}}{1972}]{1972ApJ...176....1G}
{Gunn} J.~E.,  {Gott} J.~Richard I.,  1972, \mn@doi [\apj] {10.1086/151605},
  \href {https://ui.adsabs.harvard.edu/abs/1972ApJ...176....1G} {176, 1}

\bibitem[\protect\citeauthoryear{{Hitomi Collaboration} et~al.,}{{Hitomi
  Collaboration} et~al.}{2016}]{2016Natur.535..117H}
{Hitomi Collaboration} et~al., 2016, \mn@doi [\nat] {10.1038/nature18627},
  \href {https://ui.adsabs.harvard.edu/abs/2016Natur.535..117H} {535, 117}

\bibitem[\protect\citeauthoryear{{Ho}}{{Ho}}{2008}]{2008ARA&A..46..475H}
{Ho} L.~C.,  2008, \mn@doi [\araa] {10.1146/annurev.astro.45.051806.110546},
  \href {https://ui.adsabs.harvard.edu/abs/2008ARA&A..46..475H} {46, 475}

\bibitem[\protect\citeauthoryear{{J{\'a}chym}, {Combes}, {Cortese}, {Sun}  \&
  {Kenney}}{{J{\'a}chym} et~al.}{2014}]{2014ApJ...792...11J}
{J{\'a}chym} P.,  {Combes} F.,  {Cortese} L.,  {Sun} M.,   {Kenney} J. D.~P.,
  2014, \mn@doi [\apj] {10.1088/0004-637X/792/1/11}, \href
  {https://ui.adsabs.harvard.edu/abs/2014ApJ...792...11J} {792, 11}

\bibitem[\protect\citeauthoryear{{Kanjilal}, {Dutta}  \& {Sharma}}{{Kanjilal}
  et~al.}{2021}]{2021MNRAS.501.1143K}
{Kanjilal} V.,  {Dutta} A.,   {Sharma} P.,  2021, \mn@doi [\mnras]
  {10.1093/mnras/staa3610}, \href
  {https://ui.adsabs.harvard.edu/abs/2021MNRAS.501.1143K} {501, 1143}

\bibitem[\protect\citeauthoryear{{Kauffmann} et~al.,}{{Kauffmann}
  et~al.}{2003}]{2003MNRAS.346.1055K}
{Kauffmann} G.,  et~al., 2003, \mn@doi [\mnras]
  {10.1111/j.1365-2966.2003.07154.x}, \href
  {https://ui.adsabs.harvard.edu/abs/2003MNRAS.346.1055K} {346, 1055}

\bibitem[\protect\citeauthoryear{{Kenney}, {Tal}, {Crowl}, {Feldmeier}  \&
  {Jacoby}}{{Kenney} et~al.}{2008}]{2008ApJ...687L..69K}
{Kenney} J. D.~P.,  {Tal} T.,  {Crowl} H.~H.,  {Feldmeier} J.,   {Jacoby}
  G.~H.,  2008, \mn@doi [\apjl] {10.1086/593300}, \href
  {https://ui.adsabs.harvard.edu/abs/2008ApJ...687L..69K} {687, L69}

\bibitem[\protect\citeauthoryear{{Kennicutt} \& {Evans}}{{Kennicutt} \&
  {Evans}}{2012}]{2012ARA&A..50..531K}
{Kennicutt} R.~C.,  {Evans} N.~J.,  2012, \mn@doi [\araa]
  {10.1146/annurev-astro-081811-125610}, \href
  {https://ui.adsabs.harvard.edu/abs/2012ARA&A..50..531K} {50, 531}

\bibitem[\protect\citeauthoryear{{Kent} et~al.,}{{Kent}
  et~al.}{2007}]{2007ApJ...665L..15K}
{Kent} B.~R.,  et~al., 2007, \mn@doi [\apjl] {10.1086/521100}, \href
  {https://ui.adsabs.harvard.edu/abs/2007ApJ...665L..15K} {665, L15}

\bibitem[\protect\citeauthoryear{{Kewley} \& {Dopita}}{{Kewley} \&
  {Dopita}}{2002}]{2002ApJS..142...35K}
{Kewley} L.~J.,  {Dopita} M.~A.,  2002, \mn@doi [\apjs] {10.1086/341326}, \href
  {https://ui.adsabs.harvard.edu/abs/2002ApJS..142...35K} {142, 35}

\bibitem[\protect\citeauthoryear{{Kewley}, {Dopita}, {Sutherland}, {Heisler}
  \& {Trevena}}{{Kewley} et~al.}{2001}]{2001ApJ...556..121K}
{Kewley} L.~J.,  {Dopita} M.~A.,  {Sutherland} R.~S.,  {Heisler} C.~A.,
  {Trevena} J.,  2001, \mn@doi [\apj] {10.1086/321545}, \href
  {https://ui.adsabs.harvard.edu/abs/2001ApJ...556..121K} {556, 121}

\bibitem[\protect\citeauthoryear{{Kewley}, {Nicholls}  \&
  {Sutherland}}{{Kewley} et~al.}{2019}]{2019ARA&A..57..511K}
{Kewley} L.~J.,  {Nicholls} D.~C.,   {Sutherland} R.~S.,  2019, \mn@doi [\araa]
  {10.1146/annurev-astro-081817-051832}, \href
  {https://ui.adsabs.harvard.edu/abs/2019ARA&A..57..511K} {57, 511}

\bibitem[\protect\citeauthoryear{{Khatri} \& {Gaspari}}{{Khatri} \&
  {Gaspari}}{2016}]{2016MNRAS.463..655K}
{Khatri} R.,  {Gaspari} M.,  2016, \mn@doi [\mnras] {10.1093/mnras/stw2027},
  \href {https://ui.adsabs.harvard.edu/abs/2016MNRAS.463..655K} {463, 655}

\bibitem[\protect\citeauthoryear{{Kochanek} et~al.,}{{Kochanek}
  et~al.}{2001}]{2001ApJ...560..566K}
{Kochanek} C.~S.,  et~al., 2001, \mn@doi [\apj] {10.1086/322488}, \href
  {https://ui.adsabs.harvard.edu/abs/2001ApJ...560..566K} {560, 566}

\bibitem[\protect\citeauthoryear{{Lang}, {Hogg}, {Mierle}, {Blanton}  \&
  {Roweis}}{{Lang} et~al.}{2012}]{2012ascl.soft08001L}
{Lang} D.,  {Hogg} D.~W.,  {Mierle} K.,  {Blanton} M.,   {Roweis} S.,  2012,
  {Astrometry.net: Astrometric calibration of images}

\bibitem[\protect\citeauthoryear{{Lau}, {Kravtsov}  \& {Nagai}}{{Lau}
  et~al.}{2009}]{2009ApJ...705.1129L}
{Lau} E.~T.,  {Kravtsov} A.~V.,   {Nagai} D.,  2009, \mn@doi [\apj]
  {10.1088/0004-637X/705/2/1129}, \href
  {https://ui.adsabs.harvard.edu/abs/2009ApJ...705.1129L} {705, 1129}

\bibitem[\protect\citeauthoryear{{Markevitch} \& {Vikhlinin}}{{Markevitch} \&
  {Vikhlinin}}{2007}]{2007PhR...443....1M}
{Markevitch} M.,  {Vikhlinin} A.,  2007, \mn@doi [\physrep]
  {10.1016/j.physrep.2007.01.001}, \href
  {https://ui.adsabs.harvard.edu/abs/2007PhR...443....1M} {443, 1}

\bibitem[\protect\citeauthoryear{{Mateos} et~al.,}{{Mateos}
  et~al.}{2008}]{2008A&A...492...51M}
{Mateos} S.,  et~al., 2008, \mn@doi [\aap] {10.1051/0004-6361:200810004}, \href
  {https://ui.adsabs.harvard.edu/abs/2008A&A...492...51M} {492, 51}

\bibitem[\protect\citeauthoryear{{McKee} \& {Begelman}}{{McKee} \&
  {Begelman}}{1990}]{1990ApJ...358..392M}
{McKee} C.~F.,  {Begelman} M.~C.,  1990, \mn@doi [\apj] {10.1086/168995}, \href
  {https://ui.adsabs.harvard.edu/abs/1990ApJ...358..392M} {358, 392}

\bibitem[\protect\citeauthoryear{{Merluzzi} et~al.,}{{Merluzzi}
  et~al.}{2013}]{2013MNRAS.429.1747M}
{Merluzzi} P.,  et~al., 2013, \mn@doi [\mnras] {10.1093/mnras/sts466}, \href
  {https://ui.adsabs.harvard.edu/abs/2013MNRAS.429.1747M} {429, 1747}

\bibitem[\protect\citeauthoryear{{Morandi}, {Sun}, {Mulchaey}, {Nagai}  \&
  {Bonamente}}{{Morandi} et~al.}{2017}]{2017MNRAS.469.2423M}
{Morandi} A.,  {Sun} M.,  {Mulchaey} J.,  {Nagai} D.,   {Bonamente} M.,  2017,
  \mn@doi [\mnras] {10.1093/mnras/stx1031}, \href
  {https://ui.adsabs.harvard.edu/abs/2017MNRAS.469.2423M} {469, 2423}

\bibitem[\protect\citeauthoryear{{M{\"u}ller} et~al.,}{{M{\"u}ller}
  et~al.}{2021}]{2021NatAs...5..159M}
{M{\"u}ller} A.,  et~al., 2021, \mn@doi [Nature Astronomy]
  {10.1038/s41550-020-01234-7}, \href
  {https://ui.adsabs.harvard.edu/abs/2021NatAs...5..159M} {5, 159}

\bibitem[\protect\citeauthoryear{{Nagai} \& {Lau}}{{Nagai} \&
  {Lau}}{2011}]{2011ApJ...731L..10N}
{Nagai} D.,  {Lau} E.~T.,  2011, \mn@doi [\apjl] {10.1088/2041-8205/731/1/L10},
  \href {https://ui.adsabs.harvard.edu/abs/2011ApJ...731L..10N} {731, L10}

\bibitem[\protect\citeauthoryear{{Nulsen}}{{Nulsen}}{1982}]{1982MNRAS.198.1007N}
{Nulsen} P.~E.~J.,  1982, \mn@doi [\mnras] {10.1093/mnras/198.4.1007}, \href
  {https://ui.adsabs.harvard.edu/abs/1982MNRAS.198.1007N} {198, 1007}

\bibitem[\protect\citeauthoryear{{Poggianti} et~al.,}{{Poggianti}
  et~al.}{2016}]{2016AJ....151...78P}
{Poggianti} B.~M.,  et~al., 2016, \mn@doi [\aj] {10.3847/0004-6256/151/3/78},
  \href {https://ui.adsabs.harvard.edu/abs/2016AJ....151...78P} {151, 78}

\bibitem[\protect\citeauthoryear{{Quilis}, {Moore}  \& {Bower}}{{Quilis}
  et~al.}{2000}]{2000Sci...288.1617Q}
{Quilis} V.,  {Moore} B.,   {Bower} R.,  2000, \mn@doi [Science]
  {10.1126/science.288.5471.1617}, \href
  {https://ui.adsabs.harvard.edu/abs/2000Sci...288.1617Q} {288, 1617}

\bibitem[\protect\citeauthoryear{{Rebusco}, {Churazov}, {B{\"o}hringer}  \&
  {Forman}}{{Rebusco} et~al.}{2006}]{2006MNRAS.372.1840R}
{Rebusco} P.,  {Churazov} E.,  {B{\"o}hringer} H.,   {Forman} W.,  2006,
  \mn@doi [\mnras] {10.1111/j.1365-2966.2006.10977.x}, \href
  {https://ui.adsabs.harvard.edu/abs/2006MNRAS.372.1840R} {372, 1840}

\bibitem[\protect\citeauthoryear{{Rich}, {Kewley}  \& {Dopita}}{{Rich}
  et~al.}{2011}]{2011ApJ...734...87R}
{Rich} J.~A.,  {Kewley} L.~J.,   {Dopita} M.~A.,  2011, \mn@doi [\apj]
  {10.1088/0004-637X/734/2/87}, \href
  {https://ui.adsabs.harvard.edu/abs/2011ApJ...734...87R} {734, 87}

\bibitem[\protect\citeauthoryear{{Richards} et~al.,}{{Richards}
  et~al.}{2015}]{2015ApJS..219...39R}
{Richards} G.~T.,  et~al., 2015, \mn@doi [\apjs] {10.1088/0067-0049/219/2/39},
  \href {https://ui.adsabs.harvard.edu/abs/2015ApJS..219...39R} {219, 39}

\bibitem[\protect\citeauthoryear{{Roediger} \& {Hensler}}{{Roediger} \&
  {Hensler}}{2005}]{2005A&A...433..875R}
{Roediger} E.,  {Hensler} G.,  2005, \mn@doi [\aap]
  {10.1051/0004-6361:20042131}, \href
  {https://ui.adsabs.harvard.edu/abs/2005A&A...433..875R} {433, 875}

\bibitem[\protect\citeauthoryear{{Sand} et~al.,}{{Sand}
  et~al.}{2017}]{2017ApJ...843..134S}
{Sand} D.~J.,  et~al., 2017, \mn@doi [\apj] {10.3847/1538-4357/aa7557}, \href
  {https://ui.adsabs.harvard.edu/abs/2017ApJ...843..134S} {843, 134}

\bibitem[\protect\citeauthoryear{{Sanders}, {Fabian}, {Smith}  \&
  {Peterson}}{{Sanders} et~al.}{2010}]{2010MNRAS.402L..11S}
{Sanders} J.~S.,  {Fabian} A.~C.,  {Smith} R.~K.,   {Peterson} J.~R.,  2010,
  \mn@doi [\mnras] {10.1111/j.1745-3933.2009.00789.x}, \href
  {https://ui.adsabs.harvard.edu/abs/2010MNRAS.402L..11S} {402, L11}

\bibitem[\protect\citeauthoryear{{Sarazin}}{{Sarazin}}{1988}]{1988xrec.book.....S}
{Sarazin} C.~L.,  1988, {X-ray emission from clusters of galaxies}

\bibitem[\protect\citeauthoryear{{Schlafly} \& {Finkbeiner}}{{Schlafly} \&
  {Finkbeiner}}{2011}]{2011ApJ...737..103S}
{Schlafly} E.~F.,  {Finkbeiner} D.~P.,  2011, \mn@doi [\apj]
  {10.1088/0004-637X/737/2/103}, \href
  {https://ui.adsabs.harvard.edu/abs/2011ApJ...737..103S} {737, 103}

\bibitem[\protect\citeauthoryear{{Schlegel}, {Finkbeiner}  \&
  {Davis}}{{Schlegel} et~al.}{1998}]{1998ApJ...500..525S}
{Schlegel} D.~J.,  {Finkbeiner} D.~P.,   {Davis} M.,  1998, \mn@doi [\apj]
  {10.1086/305772}, \href
  {https://ui.adsabs.harvard.edu/abs/1998ApJ...500..525S} {500, 525}

\bibitem[\protect\citeauthoryear{{Schuecker}, {Finoguenov}, {Miniati},
  {B{\"o}hringer}  \& {Briel}}{{Schuecker} et~al.}{2004}]{2004A&A...426..387S}
{Schuecker} P.,  {Finoguenov} A.,  {Miniati} F.,  {B{\"o}hringer} H.,   {Briel}
  U.~G.,  2004, \mn@doi [\aap] {10.1051/0004-6361:20041039}, \href
  {https://ui.adsabs.harvard.edu/abs/2004A&A...426..387S} {426, 387}

\bibitem[\protect\citeauthoryear{{Schure}, {Kosenko}, {Kaastra}, {Keppens}  \&
  {Vink}}{{Schure} et~al.}{2009}]{2009A&A...508..751S}
{Schure} K.~M.,  {Kosenko} D.,  {Kaastra} J.~S.,  {Keppens} R.,   {Vink} J.,
  2009, \mn@doi [\aap] {10.1051/0004-6361/200912495}, \href
  {https://ui.adsabs.harvard.edu/abs/2009A&A...508..751S} {508, 751}

\bibitem[\protect\citeauthoryear{{Scott}, {Brinks}, {Cortese}, {Boselli}  \&
  {Bravo-Alfaro}}{{Scott} et~al.}{2018}]{2018MNRAS.475.4648S}
{Scott} T.~C.,  {Brinks} E.,  {Cortese} L.,  {Boselli} A.,   {Bravo-Alfaro} H.,
   2018, \mn@doi [\mnras] {10.1093/mnras/sty063}, \href
  {https://ui.adsabs.harvard.edu/abs/2018MNRAS.475.4648S} {475, 4648}

\bibitem[\protect\citeauthoryear{{Simionescu} et~al.,}{{Simionescu}
  et~al.}{2011}]{2011Sci...331.1576S}
{Simionescu} A.,  et~al., 2011, \mn@doi [Science] {10.1126/science.1200331},
  \href {https://ui.adsabs.harvard.edu/abs/2011Sci...331.1576S} {331, 1576}

\bibitem[\protect\citeauthoryear{{Sivanandam}, {Rieke}  \&
  {Rieke}}{{Sivanandam} et~al.}{2010}]{2010ApJ...717..147S}
{Sivanandam} S.,  {Rieke} M.~J.,   {Rieke} G.~H.,  2010, \mn@doi [\apj]
  {10.1088/0004-637X/717/1/147}, \href
  {https://ui.adsabs.harvard.edu/abs/2010ApJ...717..147S} {717, 147}

\bibitem[\protect\citeauthoryear{{Smith} et~al.,}{{Smith}
  et~al.}{2010}]{2010MNRAS.408.1417S}
{Smith} R.~J.,  et~al., 2010, \mn@doi [\mnras]
  {10.1111/j.1365-2966.2010.17253.x}, \href
  {https://ui.adsabs.harvard.edu/abs/2010MNRAS.408.1417S} {408, 1417}

\bibitem[\protect\citeauthoryear{{Soto}, {Lilly}, {Bacon}, {Richard}  \&
  {Conseil}}{{Soto} et~al.}{2016}]{2016MNRAS.458.3210S}
{Soto} K.~T.,  {Lilly} S.~J.,  {Bacon} R.,  {Richard} J.,   {Conseil} S.,
  2016, \mn@doi [\mnras] {10.1093/mnras/stw474}, \href
  {https://ui.adsabs.harvard.edu/abs/2016MNRAS.458.3210S} {458, 3210}

\bibitem[\protect\citeauthoryear{{Sparre}, {Pfrommer}  \& {Ehlert}}{{Sparre}
  et~al.}{2020}]{2020MNRAS.499.4261S}
{Sparre} M.,  {Pfrommer} C.,   {Ehlert} K.,  2020, \mn@doi [\mnras]
  {10.1093/mnras/staa3177}, \href
  {https://ui.adsabs.harvard.edu/abs/2020MNRAS.499.4261S} {499, 4261}

\bibitem[\protect\citeauthoryear{{Spitzer}}{{Spitzer}}{1962}]{1962pfig.book.....S}
{Spitzer} L.,  1962, {Physics of Fully Ionized Gases}

\bibitem[\protect\citeauthoryear{{Sun} \& {Murray}}{{Sun} \&
  {Murray}}{2002}]{2002ApJ...576..708S}
{Sun} M.,  {Murray} S.~S.,  2002, \mn@doi [\apj] {10.1086/341756}, \href
  {https://ui.adsabs.harvard.edu/abs/2002ApJ...576..708S} {576, 708}

\bibitem[\protect\citeauthoryear{{Sun}, {Jones}, {Forman}, {Vikhlinin},
  {Donahue}  \& {Voit}}{{Sun} et~al.}{2007a}]{2007ApJ...657..197S}
{Sun} M.,  {Jones} C.,  {Forman} W.,  {Vikhlinin} A.,  {Donahue} M.,   {Voit}
  M.,  2007a, \mn@doi [\apj] {10.1086/510895}, \href
  {https://ui.adsabs.harvard.edu/abs/2007ApJ...657..197S} {657, 197}

\bibitem[\protect\citeauthoryear{{Sun}, {Donahue}  \& {Voit}}{{Sun}
  et~al.}{2007b}]{2007ApJ...671..190S}
{Sun} M.,  {Donahue} M.,   {Voit} G.~M.,  2007b, \mn@doi [\apj]
  {10.1086/522690}, \href
  {https://ui.adsabs.harvard.edu/abs/2007ApJ...671..190S} {671, 190}

\bibitem[\protect\citeauthoryear{{Sun}, {Voit}, {Donahue}, {Jones}, {Forman}
  \& {Vikhlinin}}{{Sun} et~al.}{2009}]{2009ApJ...693.1142S}
{Sun} M.,  {Voit} G.~M.,  {Donahue} M.,  {Jones} C.,  {Forman} W.,
  {Vikhlinin} A.,  2009, \mn@doi [\apj] {10.1088/0004-637X/693/2/1142}, \href
  {https://ui.adsabs.harvard.edu/abs/2009ApJ...693.1142S} {693, 1142}

\bibitem[\protect\citeauthoryear{{Sun}, {Donahue}, {Roediger}, {Nulsen},
  {Voit}, {Sarazin}, {Forman}  \& {Jones}}{{Sun}
  et~al.}{2010}]{2010ApJ...708..946S}
{Sun} M.,  {Donahue} M.,  {Roediger} E.,  {Nulsen} P.~E.~J.,  {Voit} G.~M.,
  {Sarazin} C.,  {Forman} W.,   {Jones} C.,  2010, \mn@doi [\apj]
  {10.1088/0004-637X/708/2/946}, \href
  {https://ui.adsabs.harvard.edu/abs/2010ApJ...708..946S} {708, 946}

\bibitem[\protect\citeauthoryear{{Sun} et~al.,}{{Sun}
  et~al.}{2021}]{2021arXiv210309205S}
{Sun} M.,  et~al., 2021, arXiv e-prints, \href
  {https://ui.adsabs.harvard.edu/abs/2021arXiv210309205S} {p. arXiv:2103.09205}

\bibitem[\protect\citeauthoryear{{Taylor}, {Davies}, {Auld}  \&
  {Minchin}}{{Taylor} et~al.}{2012}]{2012MNRAS.423..787T}
{Taylor} R.,  {Davies} J.~I.,  {Auld} R.,   {Minchin} R.~F.,  2012, \mn@doi
  [\mnras] {10.1111/j.1365-2966.2012.20914.x}, \href
  {https://ui.adsabs.harvard.edu/abs/2012MNRAS.423..787T} {423, 787}

\bibitem[\protect\citeauthoryear{{Taylor}, {Davies}, {J{\'a}chym}, {Keenan},
  {Minchin}, {Palou{\v{s}}}, {Smith}  \& {W{\"u}nsch}}{{Taylor}
  et~al.}{2016}]{2016MNRAS.461.3001T}
{Taylor} R.,  {Davies} J.~I.,  {J{\'a}chym} P.,  {Keenan} O.,  {Minchin} R.~F.,
   {Palou{\v{s}}} J.,  {Smith} R.,   {W{\"u}nsch} R.,  2016, \mn@doi [\mnras]
  {10.1093/mnras/stw1475}, \href
  {https://ui.adsabs.harvard.edu/abs/2016MNRAS.461.3001T} {461, 3001}

\bibitem[\protect\citeauthoryear{{Vazza}, {Roediger}  \& {Br{\"u}ggen}}{{Vazza}
  et~al.}{2012}]{2012A&A...544A.103V}
{Vazza} F.,  {Roediger} E.,   {Br{\"u}ggen} M.,  2012, \mn@doi [\aap]
  {10.1051/0004-6361/201118688}, \href
  {https://ui.adsabs.harvard.edu/abs/2012A&A...544A.103V} {544, A103}

\bibitem[\protect\citeauthoryear{{Vazza}, {Eckert}, {Simionescu}, {Br{\"u}ggen}
   \& {Ettori}}{{Vazza} et~al.}{2013}]{2013MNRAS.429..799V}
{Vazza} F.,  {Eckert} D.,  {Simionescu} A.,  {Br{\"u}ggen} M.,   {Ettori} S.,
  2013, \mn@doi [\mnras] {10.1093/mnras/sts375}, \href
  {https://ui.adsabs.harvard.edu/abs/2013MNRAS.429..799V} {429, 799}

\bibitem[\protect\citeauthoryear{{Voit} \& {Donahue}}{{Voit} \&
  {Donahue}}{1990}]{1990ApJ...360L..15V}
{Voit} G.~M.,  {Donahue} M.,  1990, \mn@doi [\apjl] {10.1086/185801}, \href
  {https://ui.adsabs.harvard.edu/abs/1990ApJ...360L..15V} {360, L15}

\bibitem[\protect\citeauthoryear{{Walker}, {Fabian}, {Sanders}, {Simionescu}
  \& {Tawara}}{{Walker} et~al.}{2013}]{2013MNRAS.432..554W}
{Walker} S.~A.,  {Fabian} A.~C.,  {Sanders} J.~S.,  {Simionescu} A.,   {Tawara}
  Y.,  2013, \mn@doi [\mnras] {10.1093/mnras/stt497}, \href
  {https://ui.adsabs.harvard.edu/abs/2013MNRAS.432..554W} {432, 554}

\bibitem[\protect\citeauthoryear{{Weilbacher}, {Streicher}, {Urrutia}, {Jarno},
  {P{\'e}contal-Rousset}, {Bacon}  \& {B{\"o}hm}}{{Weilbacher}
  et~al.}{2012}]{2012SPIE.8451E..0BW}
{Weilbacher} P.~M.,  {Streicher} O.,  {Urrutia} T.,  {Jarno} A.,
  {P{\'e}contal-Rousset} A.,  {Bacon} R.,   {B{\"o}hm} P.,  2012, in
  {Radziwill} N.~M.,  {Chiozzi} G.,  eds,  Society of Photo-Optical
  Instrumentation Engineers (SPIE) Conference Series Vol. 8451, Software and
  Cyberinfrastructure for Astronomy II. p. 84510B, \mn@doi{10.1117/12.925114}

\bibitem[\protect\citeauthoryear{{Weilbacher} et~al.,}{{Weilbacher}
  et~al.}{2020}]{2020A&A...641A..28W}
{Weilbacher} P.~M.,  et~al., 2020, \mn@doi [\aap]
  {10.1051/0004-6361/202037855}, \href
  {https://ui.adsabs.harvard.edu/abs/2020A&A...641A..28W} {641, A28}

\bibitem[\protect\citeauthoryear{{Willingale}, {Starling}, {Beardmore},
  {Tanvir}  \& {O'Brien}}{{Willingale} et~al.}{2013}]{2013MNRAS.431..394W}
{Willingale} R.,  {Starling} R.~L.~C.,  {Beardmore} A.~P.,  {Tanvir} N.~R.,
  {O'Brien} P.~T.,  2013, \mn@doi [\mnras] {10.1093/mnras/stt175}, \href
  {https://ui.adsabs.harvard.edu/abs/2013MNRAS.431..394W} {431, 394}

\bibitem[\protect\citeauthoryear{{Yagi}, {Komiyama}, {Yoshida}, {Furusawa},
  {Kashikawa}, {Koyama}  \& {Okamura}}{{Yagi}
  et~al.}{2007}]{2007ApJ...660.1209Y}
{Yagi} M.,  {Komiyama} Y.,  {Yoshida} M.,  {Furusawa} H.,  {Kashikawa} N.,
  {Koyama} Y.,   {Okamura} S.,  2007, \mn@doi [\apj] {10.1086/512359}, \href
  {https://ui.adsabs.harvard.edu/abs/2007ApJ...660.1209Y} {660, 1209}

\bibitem[\protect\citeauthoryear{{Yagi} et~al.,}{{Yagi}
  et~al.}{2010}]{2010AJ....140.1814Y}
{Yagi} M.,  et~al., 2010, \mn@doi [\aj] {10.1088/0004-6256/140/6/1814}, \href
  {https://ui.adsabs.harvard.edu/abs/2010AJ....140.1814Y} {140, 1814}

\bibitem[\protect\citeauthoryear{{Yagi}, {Gu}, {Fujita}, {Nakazawa}, {Akahori},
  {Hattori}, {Yoshida}  \& {Makishima}}{{Yagi}
  et~al.}{2013}]{2013ApJ...778...91Y}
{Yagi} M.,  {Gu} L.,  {Fujita} Y.,  {Nakazawa} K.,  {Akahori} T.,  {Hattori}
  T.,  {Yoshida} M.,   {Makishima} K.,  2013, \mn@doi [\apj]
  {10.1088/0004-637X/778/2/91}, \href
  {https://ui.adsabs.harvard.edu/abs/2013ApJ...778...91Y} {778, 91}

\bibitem[\protect\citeauthoryear{{Yagi}, {Gu}, {Koyama}, {Nakata}, {Kodama},
  {Hattori}  \& {Yoshida}}{{Yagi} et~al.}{2015}]{2015AJ....149...36Y}
{Yagi} M.,  {Gu} L.,  {Koyama} Y.,  {Nakata} F.,  {Kodama} T.,  {Hattori} T.,
  {Yoshida} M.,  2015, \mn@doi [\aj] {10.1088/0004-6256/149/2/36}, \href
  {https://ui.adsabs.harvard.edu/abs/2015AJ....149...36Y} {149, 36}

\bibitem[\protect\citeauthoryear{{Yagi}, {Yoshida}, {Gavazzi}, {Komiyama},
  {Kashikawa}  \& {Okamura}}{{Yagi} et~al.}{2017}]{2017ApJ...839...65Y}
{Yagi} M.,  {Yoshida} M.,  {Gavazzi} G.,  {Komiyama} Y.,  {Kashikawa} N.,
  {Okamura} S.,  2017, \mn@doi [\apj] {10.3847/1538-4357/aa68e3}, \href
  {https://ui.adsabs.harvard.edu/abs/2017ApJ...839...65Y} {839, 65}

\bibitem[\protect\citeauthoryear{{Yan} \& {Blanton}}{{Yan} \&
  {Blanton}}{2012}]{2012ApJ...747...61Y}
{Yan} R.,  {Blanton} M.~R.,  2012, \mn@doi [\apj] {10.1088/0004-637X/747/1/61},
  \href {https://ui.adsabs.harvard.edu/abs/2012ApJ...747...61Y} {747, 61}

\bibitem[\protect\citeauthoryear{{Yoshida} et~al.,}{{Yoshida}
  et~al.}{2002}]{2002ApJ...567..118Y}
{Yoshida} M.,  et~al., 2002, \mn@doi [\apj] {10.1086/338353}, \href
  {https://ui.adsabs.harvard.edu/abs/2002ApJ...567..118Y} {567, 118}

\bibitem[\protect\citeauthoryear{{Yoshida} et~al.,}{{Yoshida}
  et~al.}{2008}]{2008ApJ...688..918Y}
{Yoshida} M.,  et~al., 2008, \mn@doi [\apj] {10.1086/592430}, \href
  {https://ui.adsabs.harvard.edu/abs/2008ApJ...688..918Y} {688, 918}

\bibitem[\protect\citeauthoryear{{Yoshida}, {Yagi}, {Komiyama}, {Furusawa},
  {Kashikawa}, {Hattori}  \& {Okamura}}{{Yoshida}
  et~al.}{2012}]{2012ApJ...749...43Y}
{Yoshida} M.,  {Yagi} M.,  {Komiyama} Y.,  {Furusawa} H.,  {Kashikawa} N.,
  {Hattori} T.,   {Okamura} S.,  2012, \mn@doi [\apj]
  {10.1088/0004-637X/749/1/43}, \href
  {https://ui.adsabs.harvard.edu/abs/2012ApJ...749...43Y} {749, 43}

\bibitem[\protect\citeauthoryear{{Zhang} et~al.,}{{Zhang}
  et~al.}{2013}]{2013ApJ...777..122Z}
{Zhang} B.,  et~al., 2013, \mn@doi [\apj] {10.1088/0004-637X/777/2/122}, \href
  {https://ui.adsabs.harvard.edu/abs/2013ApJ...777..122Z} {777, 122}

\bibitem[\protect\citeauthoryear{{Zhuravleva} et~al.,}{{Zhuravleva}
  et~al.}{2014}]{2014Natur.515...85Z}
{Zhuravleva} I.,  et~al., 2014, \mn@doi [\nat] {10.1038/nature13830}, \href
  {https://ui.adsabs.harvard.edu/abs/2014Natur.515...85Z} {515, 85}

\makeatother
\end{thebibliography}
\end{document}